\documentclass[12pt]{cernart}
\usepackage{epsfig}
\usepackage{amsmath}
\usepackage{amssymb}
\usepackage{times}
\usepackage{multirow}

\begin{document}
 
\setcounter{topnumber}{3}
\renewcommand{\topfraction}{0.999}
\renewcommand{\bottomfraction}{0.99}
\renewcommand{\textfraction}{0.0}
\setcounter{totalnumber}{6}
 
\def\thefootnote{\fnsymbol{footnote}}
\begin{titlepage}
%
\title{\large{Charged Pion Production in p+C Collisions \\
              at 158 GeV/c Beam Momentum: Discussion}}

%

\begin{Authlist}
\vspace{2mm}
\noindent
G.~Barr$^{5}$, O.~Chvala$^{6}$, H.G.~Fischer$^{4}$, M.~Kreps$^{1}$,
M.~Makariev$^{7}$, C.~Pattison$^{5}$, A.~Rybicki$^{3}$, 
D.~Varga$^{4,2}$, 
S.~Wenig$^{9,}$\footnote{Corresponding author: Siegfried.Wenig@cern.ch}\\
\vspace*{5mm}
\noindent
$^{1}$ Comenius University, Bratislava, Slovakia \\
$^{2}$ E\"otv\"os Lor\'ant University, Budapest, Hungary \\
$^{3}$ The H. Niewodnicza\'nski Institute of Nuclear Physics, 
       Polish Academy of Sciences, Cracow, Poland \\
$^{4}$ CERN, Geneva, Switzerland \\
$^{5}$ Oxford University, Oxford, UK \\
$^{6}$ Institute of Particle and Nuclear Physics, Charles University, 
       Prague, Czech Republic \\
$^{7}$ Atomic Physics Department, Sofia University St. Kliment Ohridski, 
       Sofia, Bulgaria \\
\end{Authlist}
\vspace{5cm}

\begin{abstract}
\vspace{-3mm}
The new data on pion production in p+C interactions from the NA49 
experiment at the CERN SPS are used for a detailed study of hadronization
in the collision of protons with light nuclei. The comparison to the
extensive set of data from p+p collisions obtained with the same detector 
allows for the separation and extraction of the projectile
and target contributions to the pion yield both in longitudinal and in
transverse momentum.
\end{abstract}

\cleardoublepage
\end{titlepage}

%
%
\section{Introduction}
\vspace{3mm}
\label{sec:intro}

Hadron-nucleus interactions constitute a unique laboratory for the study
of hadronization in the non-perturbative sector of QCD by offering, in
particular, access to multiple hadronic collision processes and by
opening an important link to nucleus-nucleus reactions. In fact the
latter process cannot be understood properly without reference to the
more elementary hadron-hadron and hadron-nucleus interactions. Past
attempts in this field have been largely frustrated by the absence of 
detailed and consistent data sets both from hadron-hadron and from 
hadron-nucleus collisions. The NA49 experiment aims at providing such
data sets obtained with the same detector over a wide range of projectile
and target combinations \cite{bib:hgf} by measuring yields of identified
secondaries with a dense coverage of the available phase space.

A recent publication concerning pion production in p+p collisions
\cite{bib:pp_paper} has now been followed by data of similar quality 
from minimum-bias p+C interactions \cite{bib:pc_paper}. This allows for 
a new attempt at comparing both
reactions and extracting quantitative information about hadronization
involving light nuclei. The hadronization process in hadron-nucleus
interactions consists of three distinct components:
 
\begin{itemize}
\item The projectile fragmentation which carries the imprint of multiple
      collisions in nuclear matter and is the part of principle interest
      for this study.
\item The fragmentation of the participating target nucleons. 
\item The contribution from intranuclear cascading which is generated
      by the propagation and interaction of the participating nucleons 
      and produced hadrons inside the nucleus.
\end{itemize}

The separation of the three components is non-trivial if it is to be 
conducted in a model-independent way. It has been demonstrated by
NA49 for the first time in p+Pb collisions concerning net proton
production \cite{bib:hgf}. In this case the full range of projectile and 
target combinations available to the experiment has been used, with
baryon number conservation as the only external constraint.

For pions the situation is more complex as there is no conservation
law (except charge conservation) to be exploited. In addition the 
target cascading part extends closer to the central region than for
protons, up to $x_F$ values of about -0.05. Both the target and
the projectile components feed-over into the respective opposite 
hemispheres with a priori unknown ranges and shapes of the 
longitudinal momentum distributions. These parameters should be 
determined experimentally without taking recourse to unfounded model
assumptions.

The present study will attempt the quantification and separation
of the relevant components, with a special view to the extraction
of the projectile contribution in its longitudinal and transverse
momentum behaviour. For this aim the reference to the elementary
hadron-proton interactions is absolutely essential and most of
the argumentation will be established in direct comparison to these 
processes, invoking isospin symmetry as an important constraint 
for the constitution of the target contribution.

In its first part the paper deals with $p_T$ integrated $x_F$ 
distributions since the superposition process is to first order a 
matter of longitudinal momentum space. After a presentation of the 
longitudinal pion yields in relation to p+p collisions in 
Sect.~\ref{sec:comp_pp}, constraints on the overlap of target and 
projectile components from elementary interactions are developed 
in Sect.~\ref{sec:pp_two}, resulting in a two-component picture 
of these collisions. This two-component picture is applied to the p+C 
reaction in Sect.~\ref{sec:pc_two} which provides detailed 
information about the target and projectile fragmentation. Nuclear 
aspects are discussed in Sect.~\ref{sec:aspects}, followed by the 
quantification of the cascading contribution in Sect.~\ref{sec:casc}. 
This allows for the extraction of the multiple collision part of the 
interaction for a number of typical observables, and a comparison to 
the centrality-constrained data available 
from the experiment in Sect.~\ref{sec:mult_coll}.

In its second part the paper will exploit the more detailed behaviour
of double inclusive cross sections which give access to the study of
transverse momentum behaviour. Sect.~\ref{sec:pt_dist} presents the general 
$p_T$ dependencies in relation to p+p interactions and Sect.~\ref{sec:high_pt}
addresses in particular the behaviour towards higher $p_T$ up to about
1.8~GeV/c. A discussion of the mass and $p_T$ dependence of the 
feed-over mechanism in relation to resonance decay will close the paper
in Sect.~10.

%
%
\part{$\mathbf{p_T}$ Integrated $\mathbf{x_F}$ Distributions}
\vspace{3mm}
\label{part:one}
%
%
\section{Comparison to p+p interactions}
\vspace{3mm}
\label{sec:comp_pp}

%
%
\subsection{Charge-average $\mathbf{x_F}$ distributions}
\vspace{3mm}

\begin{figure}[h]
  \centering
  \epsfig{file=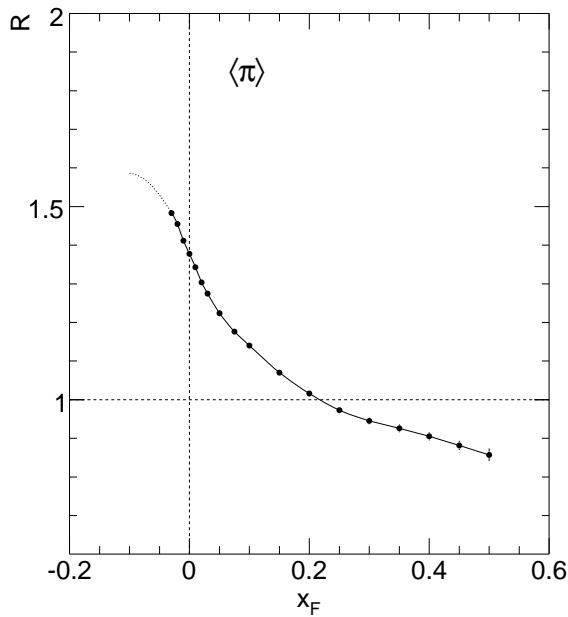,width=7.5cm}
  \caption{Ratio $R(x_F)$ as a function of $x_F$ for the charge-average pion
           $\langle \pi \rangle$}
  \label{fig:c2p_mean}
\end{figure}

The most general reduction of the multiparticle phase space of
pion production that may be obtained is the $p_T$ integrated, 
charge-averaged ($\langle \pi \rangle = ( \pi^+ + \pi^- )/2$) longitudinal 
momentum distribution, in the following characterized by the scaling variable

\begin{equation}
  x_F = \frac{p_L}{\sqrt{s}/2}
\end{equation}
where $p_L$ and $\sqrt{s}$ are given in the nucleon-nucleon cms. This 
distribution is compared to the p+p interaction \cite{bib:pp_paper} 
by forming the ratio

\begin{equation}
  R(x_F) = \frac{(dn/dx_F)^{\textrm{pC}}}
  {(dn/dx_F)^{\textrm{pp}}}
  \label{eq:yield_rat}
\end{equation}
and shown in Fig.~\ref{fig:c2p_mean}. Here the integrated pion densities 
$dn/dx_F$ are obtained from the interpolated double differential cross 
sections as described in detail in \cite{bib:pp_paper} and \cite{bib:pc_paper}.

Since the acceptance for particle identification of the NA49 detector
imposes a limit \cite{bib:pc_paper} at $x_F$~=~-0.03 for the complete 
integration of the $p_T$ distributions, a slight extrapolation to 
$x_F\textrm{ = -0.1}$ is also indicated in Fig.~\ref{fig:c2p_mean}. This 
extrapolation uses an extension of the data interpolation developed 
for the p+C data \cite{bib:pc_paper} into the non-measured 
region of $p_T$. It will be justified in the following sections of this paper. 
The upper limit of the comparison at $x_F$~=~0.5 is due to the limited event 
statistics in the p+C data sample.

Several features of interest emerge from this first comparison. The ratio
R decreases smoothly from about 1.6 at the limit of $x_F$~=~-0.1 via
1.38 at $x_F$~=~0 to a value of 0.85 at $x_F$~=~0.5. The factor 1.6 is 
close to the expected mean number of projectile collisions 
$\langle \nu \rangle$ inside the Carbon nucleus (see Sect.~\ref{sec:aspects}) 
and the value 1.38 at $x_F$~=~0 is close to the expectation 
($\langle \nu \rangle$~+~1)/2 from the most straight-forward 
projectile-target superposition scheme \cite{bib:num_coll}. The decrease 
of particle density in the far forward direction recalls a known 
phenomenon of projectile fragmentation. It corresponds to the $x_F$ 
dependence of the exponent $\alpha$ in the parametrization

\begin{equation}
  \frac{d\sigma}{dx_F} \sim A^{\alpha}
\end{equation}
where A is the nuclear weight. For a limited range in transverse
momentum and for the far forward direction alpha has been shown
to decrease as a function of $x_F$ \cite{bib:barton}.

The increased particle density in the region -0.1~$< x_F <$~0.2 on the
other hand is due to the superposition of the two principle components
of the hadronization process, namely the target nucleons participating
in the collision and the projectile itself. The quantification and
extraction of these two basic ingredients to the p+A interaction
is the main aim of this study which can for the first time make full
use of the important transition region between the two components.
This transition region will be shown to be located between the limits
-0.1~$< x_F <$~0.1.

\begin{figure}[h]
  \centering
  \epsfig{file=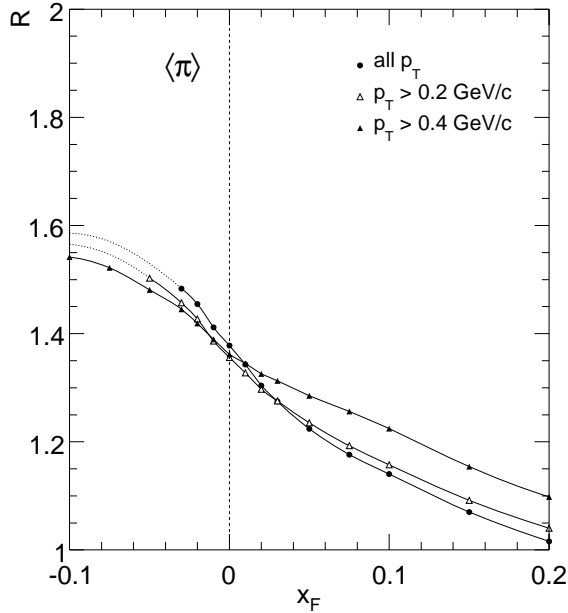,width=7.5cm}
  \caption{Ratio $R(x_F)$ as a function of $x_F$ for the charge-average pion
           $\langle \pi \rangle$ obtained with lower integration limits at
           $p_T$~= 0, 0.2 and 0.4 GeV/c}
  \label{fig:c2p_ptcut}
\end{figure}

A first indication concerning the validity of the extrapolation
towards $x_F$~=~-0.1 used in Fig.~\ref{fig:c2p_mean} is obtained by 
studying the sensitivity of the integrated yields to a lower cut-off 
in $p_T$ as shown in Fig.~\ref{fig:c2p_ptcut}.

Here the ratio $R(x_F)$ is obtained for lower integration limits at 
$p_T$~=~0, 0.2 and 0.4 GeV/c. For the latter value the full region down
to $x_F$~=~-0.1 is experimentally available. It appears that the effect
of the successively increasing cut-offs is on the few percent level
for negative $x_F$ and can therefore be considered as a second order
phenomenon. It finds its explanation in the $p_T$ dependence of the
range in $x_F$ of the projectile feed-over as discussed in 
Sect.~\ref{sec:high_pt}. The quantitatively larger effect in the forward 
hemisphere is noteworthy. It has its origin in the sizeable increase in 
$\langle p_T \rangle$ for p+C compared to p+p interactions with 
ultimately very important $p_T$ dependencies in the range 
$p_T >$~1.5~GeV/c (Sect.~\ref{sec:pt_dist}).

%
%
\subsection{Charge dependence}
\vspace{3mm}

\begin{figure}[b!]
  \centering
  \epsfig{file=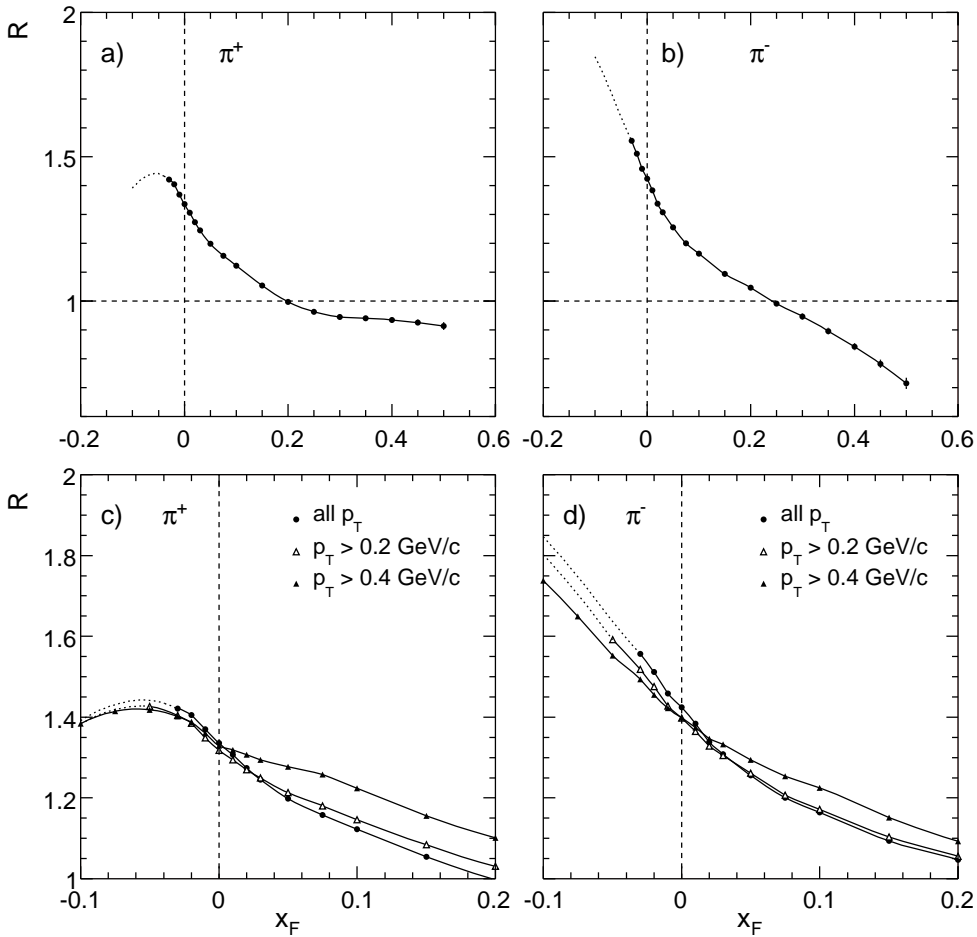,width=13cm}
  \caption{Ratio $R(x_F)$ as a function of $x_F$ without $p_T$ cut off (two
           upper panels) and with $p_T$ cut off included (two lower panels) 
	   for a) and c) $\pi^+$,  b) and d) $\pi^-$ }
  \label{fig:c2p_charge}
\end{figure}

The yield ratio $R(x_F)$ is plotted separately for $\pi^+$ and $\pi^-$ 
secondaries in Fig.~\ref{fig:c2p_charge}. Evidently there are differences 
both in the far forward, the projectile hemisphere, and more pronounced, 
in the backward region. The former effect results in a sizeable 
increase of the $\pi^+$/$\pi^-$ ratio for $x_F >$~0.3 and signals 
an important modification of the projectile fragmentation mechanism. 
The latter effect is directly related to the target. Due to isospin 
symmetry \cite{bib:hgf1} the neutron content of the Carbon nucleus enhances 
the $\pi^-$ and suppresses the $\pi^+$ yields from the target relative 
to the p+p interaction. The observation of the $\pi^+$/$\pi^-$ 
ratio in this region can therefore provide direct, model-independent 
access to the relative target contribution.

The ratio R for different lower cut-offs in the $p_T$ integration is also
shown in Fig.~\ref{fig:c2p_charge}. For $\pi^+$ a local maximum of R at 
$x_F \sim$~-0.05 is followed by
a decrease towards lower $x_F$. For $\pi^-$ on the other hand a steady increase
of R through the same region of $x_F$ is observed. These effects will be 
studied quantitatively in Sect.~\ref{sec:pc_two}.

%
%
\subsection{Charge ratios}
\vspace{3mm}

\begin{figure}[h]
  \centering
  \epsfig{file=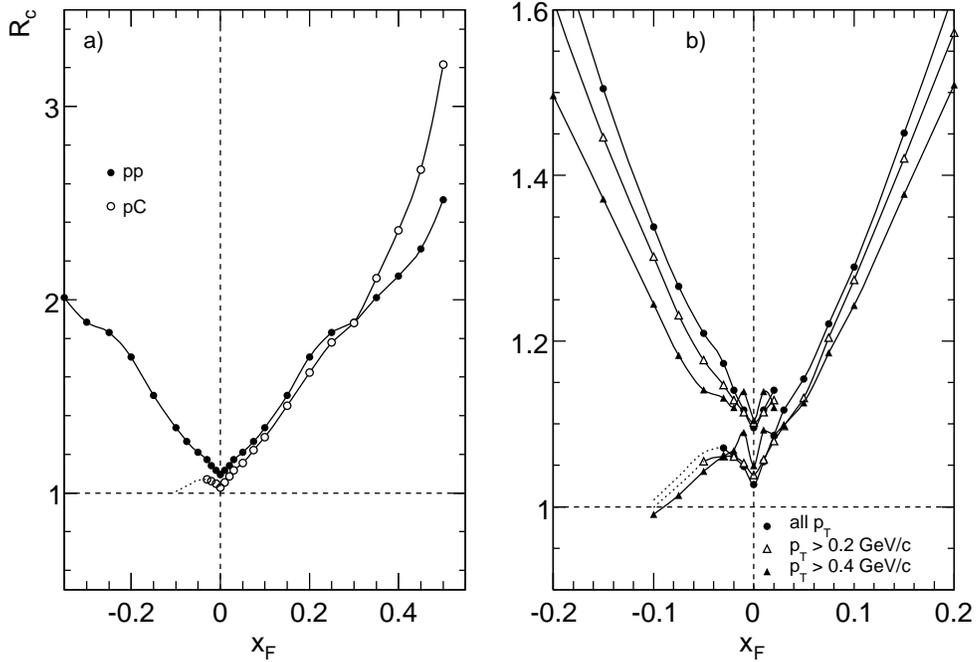,width=13cm}
  \caption{$p_T$ integrated $\pi^+$/$\pi^-$ ratio as a function of $x_F$ in
           p+p and p+C a) without $p_T$ cut off and b) with $p_T$ cut off included
           as the data points in p+p collisions are plotted only up to 
           $x_F$~=~0.02 }
  \label{fig:c2p_rat}
\end{figure}

The effects discussed above result in a strong backward-forward asymmetry
of the charge ratio

\begin{equation}
  R_c(x_F) = \frac{(dn/dx_F)^{\pi^+}(x_F)}{(dn/dx_F)^{\pi^-}(x_F)}
\end{equation}
for the p+C interaction as compared to p+p. This is demonstrated in 
Fig.~\ref{fig:c2p_rat} which shows $R_c$ for both the p+C and the p+p reaction as 
a function of $x_F$. An enlarged view of the central and backward $x_F$ regions 
is also presented in Fig.~\ref{fig:c2p_rat} for different $p_T$ cut-offs in 
the transverse momentum integration, giving further support to the extrapolation
used at low $p_T$. 

Again invoking isospin symmetry, an approach of $R_c$ to unity will indicate
the region of prevailing target fragmentation, whereas an approach to
the values measured in p+p collisions will signal the prevalence of
projectile fragmentation. Indeed the plots of Fig.~\ref{fig:c2p_rat} allow 
to argue that both expectations are fulfilled in the regions $x_F<$~-0.1 
and $x_F>$~+0.1, respectively. The situation is, however, more complex in detail.
The approach to $R_c$~=~1 is not smooth but shows structure with a local
maximum at $x_F\sim$~-0.05, and the $R_c$ values for positive $x_F$ approach but do
not quite reach the ones measured in p+p collisions below $x_F$~=~0.2. For
$x_F >$~0.3 on the other hand, they start to exceed these values. A more
detailed discussion of these second-order effects will be given in
Sect.~\ref{sec:pc_two} below.    

%
%
\section{Projectile and target contributions in elementary interactions}
\vspace{3mm}
\label{sec:pp_two}

The superposition and separation of the target and projectile components
in p+A collisions is a main ingredient for the argumentation of this paper.  
The inspection of these two components also in the elementary hadron-proton 
interactions is therefore a necessary condition for the model-independent
interpretation of the data. The discussion will be built around three
experimental facts: 
\begin{itemize}
\item The absence of long-range two-particle correlations at $|x_F| >$~0.2.
\item The presence of forward-backward multiplicity correlations at $|x_F| <$~0.1.
\item The $x_F$ dependence of the $\pi^+$/$\pi^-$ ratio in 
      $\langle \pi \rangle$+p collisions.
\end{itemize}
%
%
\subsection{Long-range two particle correlations}
\vspace{3mm}
\label{sec:fb_correl}

\begin{figure}[b!]
  \centering
  \epsfig{file=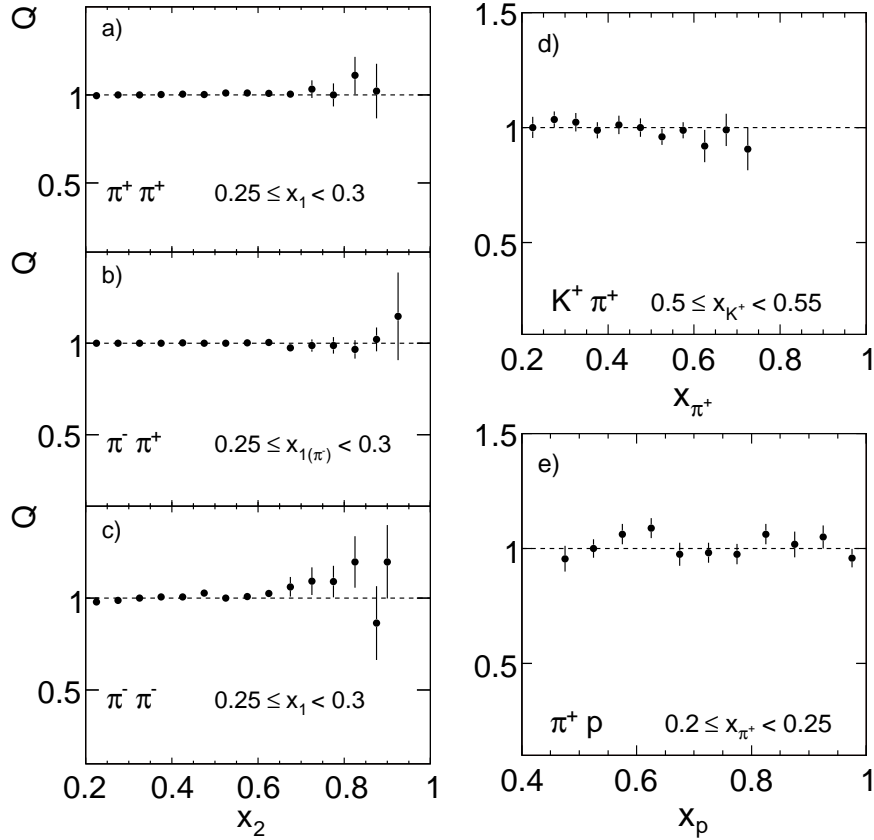,width=11.6cm}
  \caption{Forward-backward correlations in p+p collisions measured by 
           \cite{bib:bobbink} for: a) $\pi^+\pi^+$, b) $\pi^-\pi^+$, 
	   c) $\pi^-\pi^-$, d) K$^+\pi^+$, and e) $\pi^+$p }
  \label{fig:fb_correl}
\end{figure}

A high precision study of the forward-backward correlation function

\begin{equation}
  Q(x_1,x_2) = \frac{g(x_1,x_2)}{f(x_1)\cdot f(x_2)}\sigma_{\textrm{trig}}
\end{equation}
has been performed at the CERN ISR \cite{bib:bobbink} for different 
combinations of identified secondary hadrons in p+p interactions. Here 
$x_1$ and $x_2$ are the Feynman x values of the two particles which are 
selected in the opposite hemispheres of the reaction and $f$ and $g$ 
denote the invariant one and two-particle cross sections, respectively.
Typical plots of $Q(x_1,x_2)$ are shown in Fig.~\ref{fig:fb_correl} for
pion pairs and for K$\pi$ and p$\pi$ combinations.

The acceptance and particle identification constraints for the two 
spectrometer arms of the ISR experiment limit the longitudinal coverage 
to $|x_F| >$~0.2 for pions and $|x_F| >$~0.4 for protons and kaons. Within these
limits, no deviation of $Q(x_1,x_2)$ from unity is visible within tight 
error bars with the only exception of the regions of large $x_F$ where
energy-momentum conservation will finally impose some correlations
of kinematic origin \cite{bib:bobbink}. 

Some of the  important results of this experiment concern the absence
of charge or flavour exchange, the indication of gluon exchange as the
exclusive exchange mechanism in soft hadronic collisions at the SPS/ISR
energy, and the proof that no notable forward-backward correlations 
exist above the limit of $|x_F|$~=~0.2.

%
%
\subsection{Forward-backward multiplicity correlations}
\vspace{3mm}
\label{sec:fb_mult}

A series of bubble chamber experiments \cite{bib:bromberg,bib:kafka,bib:aivazyan} 
at SPS energies, and a streamer chamber experiment at the ISR 
\cite{bib:uhlig} have extended the study of \cite{bib:bobbink} to 
full phase space by establishing multiplicity correlations 
between the forward and the backward hemisphere. In experiments 
[8-10] the complete $x_F$ region is accessible
by momentum measurement in the fixed-target laboratory frame. Since
there is no particle identification available, the assumption of
pion mass for the necessary Lorentz transformation creates a certain
but practically negligible bias in the determination of central
rapidity. Experiment \cite{bib:uhlig} has no momentum measurement and only
determines pseudo-rapidity $\eta$ but the $\eta$~=~0 reference is uniquely
given for all particles by the angle measurement alone due to the 
collider geometry of the ISR. 

\begin{figure}[h]
  \centering
  \epsfig{file=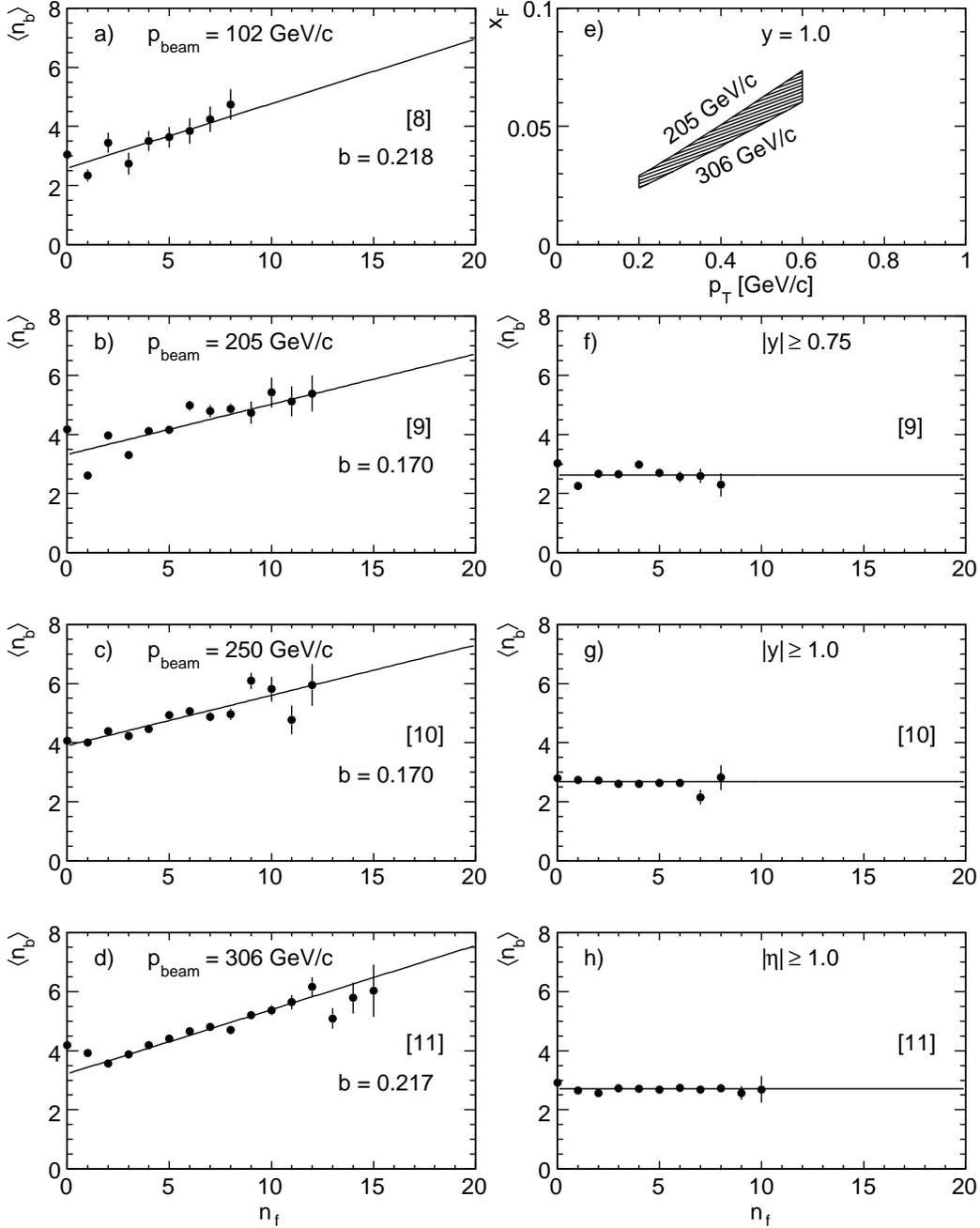,width=14cm}
  \caption{Forward-backward multiplicity correlations, a) to d) experimental
           data and fits, [8-11]; f) to h) correlation with rapidity
	   selection as given, [9-11]; d) $x_F$ as a function of $p_T$ at
	   $y$~=~1 for beam momentum between 205 and 306 GeV/c }
  \label{fig:fb_mult}
\end{figure}

Forward-backward multiplicity correlations are quantified by splitting
the charged multiplicity $n_{\textrm{ch}}$ per event into 
a forward ($n_f$) and backward ($n_b$) sample around $y$~=~0 or 
$\eta$~=~0 and by plotting the average backward multiplicity 
$\langle n_b \rangle$ as a function of $n_f$ as shown in 
Fig.~\ref{fig:fb_mult}.

Except for some odd-even fluctuations at small multiplicities, a 
linear correlation of the type

\begin{equation}
  \langle n_b \rangle = a + b \cdot n_f
\end{equation}
emerges for all experiments, with values of the slope parameter
$b$ between 0.17 and 0.22 in the $\sqrt{s}$ range from 14 to 24 GeV 
(Fig.~\ref{fig:fb_mult}a-d). These results do not contradict the 
absence of correlations shown by \cite{bib:bobbink} above. In fact the observed 
multiplicity correlations vanish if only particles above one unit of 
rapidity or pseudo-rapidity are allowed to contribute to the sample,
as presented in Fig.~\ref{fig:fb_mult}f-h. For the region of $\sqrt{s}$
covered and for transverse momenta in the neighbourhood of 
$\langle p_T \rangle$ this corresponds to a cut in $x_F$ between 
0.04 and 0.06 units as shown in Fig.~\ref{fig:fb_mult}e. This 
limit is well below the $x_F$ acceptance of \cite{bib:bobbink}.

The combined results of Sects.~\ref{sec:fb_correl} and \ref{sec:fb_mult} 
allow the statement that in the SPS energy range there is a limited 
feed-over of (mostly) pion production from the forward to the backward 
hemisphere and vice-versa. This feed-over contains between 17 and 21\% 
of the particles produced in each hemisphere and is limited to the 
region of $|x_F| \lesssim$~0.05. There is no further long-range 
correlation beyond this limit. This statement has to be seen within 
the experimental uncertainties of Fig.~\ref{fig:fb_mult}f,g, and h. 
A tail of feed-over up to $x_F \sim$~0.1 as it is following from the 
study of charge ratios, Sect.~\ref{sec:pi_rat}, will correspond to a 
percent effect on the slope parameter $b$ and is compatible with the 
error bars.

Before coming to a further quantification of this situation it is
worth while commenting on the energy dependence of the correlation
determined by \cite{bib:uhlig}. In fact the correlation factor b increases
steadily with $\sqrt{s}$ and reaches a value of $b$~=~0.31 at $\sqrt{s}$~=~62
GeV. It has been established \cite{bib:uhlig} that this increase is not accompanied 
by an extension of the feed-over range in $x_F$ by showing that the
correlation vanishes by replacing the cut at $|\eta| >$~1 by an $s$-dependent
one at $|\eta| >$~$0.5\ln{s} - 2$ . This cut happens to correspond in first 
order to an $s$-independent $x_F$ cut at 0.05. The increase in the 
correlation is therefore due to the non-scaling increase of central pion 
production characterized by the rising ``rapidity plateau'' \cite{bib:guettler}.

%
%
\subsection{ $\pi^+$/$\pi^-$ ratio in pion-induced interactions}
\vspace{3mm}
\label{sec:pi_rat}

The range and shape of the feed-over components of hadronic interactions
can also be constrained by using asymmetric collisions and 
by exploiting a measured quantity that exhibits a predicted behaviour
for either the projectile or the target fragmentation.

Such a combination is given by using the $\pi^+$/$\pi^-$ ratio for 
$\langle \pi \rangle$+p interactions. Here $\langle \pi \rangle$+p 
indicates the average of $\pi^+$+p and $\pi^-$+p
collisions. In this case, the $\pi^+$/$\pi^-$ ratio is exactly unity for
the complete projectile fragmentation. This follows in a completely
model-independent way from the fulfillment of isospin symmetry
in hadronic collisions. By plotting the $\pi^+$/$\pi^-$ ratio as a function
of $x_F$ one therefore obtains directly the extent and shape of the
proton target feed-over into the projectile hemisphere using the
deviation of the measured charge ratio from unity.

Experimentally this means the performance of precision measurements
of identified pion yields in the central region both with $\pi^+$ and $\pi^-$ 
beams. This has been achieved with the NA49 detector \cite{bib:hgf}. It has
been shown that isospin symmetry, i.e. the inversion of $\pi^+$ and $\pi^-$
yields when passing from a $\pi^+$ to a $\pi^-$ projectile, is indeed fulfilled
for $x_F >$~0.1 in this reaction \cite{bib:hgf1}. 

In the determination of the $\pi^+$/$\pi^-$ ratio all experimental corrections 
drop with the exception of the feed-down from weak decays of strange particles 
for the proton target contribution. This feed-down which is in principle 
asymmetric in its $\pi^+$ and $\pi^-$ yields has been quantified. It contributes 
less than 0.5\% to the $\pi^+$/$\pi^-$ ratio in the forward region and has been 
corrected for.

\begin{figure}[h]
  \centering
  \epsfig{file=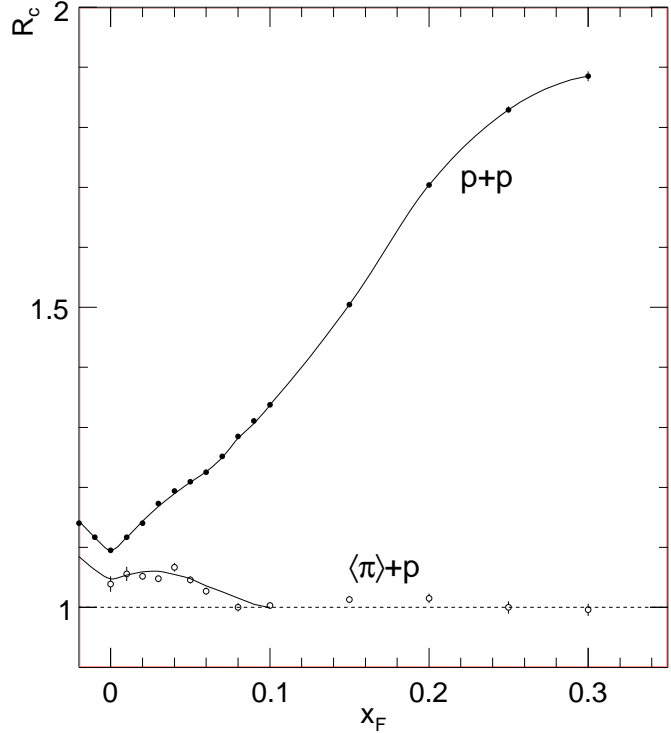,width=9cm}
  \caption{$\pi^+$/$\pi^-$ ratio in p+p and $\langle \pi \rangle$+p}
  \label{fig:pipr_rat}
\end{figure}

The resulting pionic charge ratio is shown in Fig.~\ref{fig:pipr_rat} as a 
function of $x_F$ for $\langle \pi \rangle$+p collisions in comparison 
to the same quantity in p+p interactions. 

For $\langle \pi \rangle$+p the $\pi^+$/$\pi^-$ ratio levels off to unity 
at $x_F$ above 0.08 and stays constant at this value up to the limit of data
extraction at $x_F$~=~0.3. The same ratio for p+p collisions shows the
known strong increase with $x_F$ \cite{bib:pp_paper} which is reflecting the 
presence of the proton projectile. Surprisingly, the approach of the charge
ratio to unity is not smooth but shows a characteristic structure
with a secondary maximum at $x_F \sim$~0.04 which is reminiscent of the
similar behaviour of this ratio in the target hemisphere of
p+C interactions, Fig.~\ref{fig:c2p_rat}. It should be emphasized here 
that the precision of the charge ratio as far as statistical and 
systematic errors are concerned, is on the sub-percent level in this region. 

%
%
\subsection{A two-component picture of pion production in p+p and 
            $\langle \pi \rangle$+p collisions}
\vspace{3mm}

\begin{figure}[h]
  \centering
  \epsfig{file=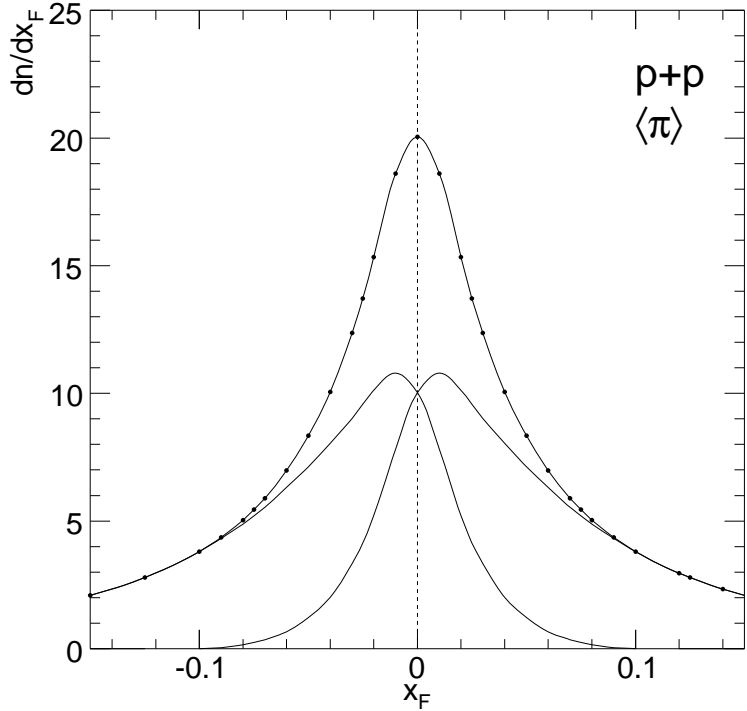,width=10cm}
  \caption{Two-component picture of charge-averaged pion production in p+p
          collisions, showing the symmetric contribution from the target
	  and projectile and their sum corresponding to the data 
	  \cite{bib:pp_paper} }
  \label{fig:pp_2comp}
\end{figure}

The combined experimental evidence collected in the preceding sections
allows for the construction of a two-component picture for pion
production in elementary hadronic interactions. The two components
correspond to the target and the projectile contributions. These
contributions obey factorization, a well known phenomenon in hadronic
reactions which specifies that the target component is independent
of the projectile particle type, and vice-versa for the projectile
fragmentation. Both components feed-over into the opposite hemisphere
with a range that has a maximum extent of 0.1 units of $x_F$ in the
SPS/ISR energy range, Sect.~\ref{sec:pi_rat}. Fixing this range, using 
the symmetry constraint at $x_F$~=~0 in p+p collisions and limiting the 
integral of particle yield in the opposite hemisphere to about 19\% of the 
total yield per component, Sect.~\ref{sec:fb_mult}, there is very limited 
freedom in the construction of this two-component picture. 
Fig.~\ref{fig:pp_2comp} shows the $x_F$ dependence of the two components 
as it is obtained obeying the above constraints, and their superposition 
to the measured total inclusive yield for p+p collisions.

\begin{figure}[h]
  \centering
  \epsfig{file=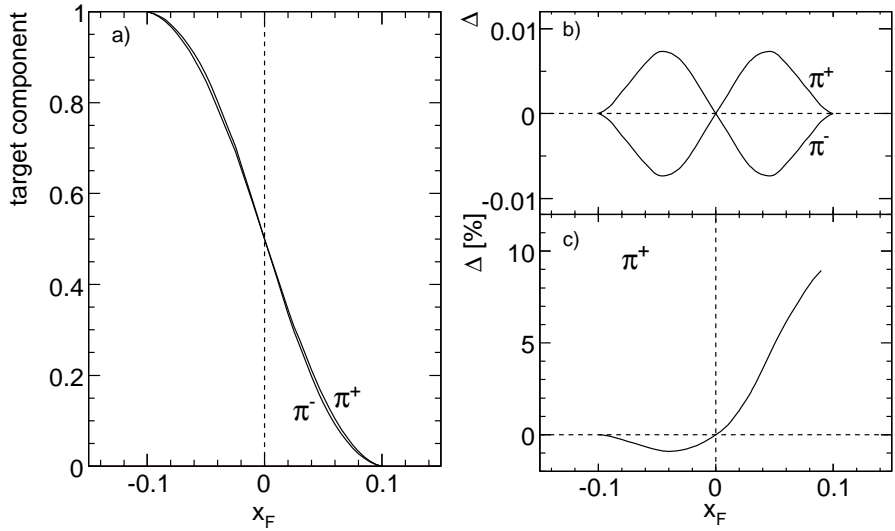,width=12cm}
  \caption{Target component of $\pi^+$ and $\pi^-$: a) absolute value with
    respect to the total yield, b) absolute deviation from 
    the target component of $\langle \pi \rangle$, and c) deviation in percent 
    for $\pi^+$ }
  \label{fig:fover_shape}
\end{figure}

The behaviour in the feed-over region -0.1~$< x_F <$~0.1 can be further
quantified by plotting the ratio of the proton target component relative
to the total inclusive pion yield

\begin{equation}
  R_{\textrm{two-comp}} = \frac{(dn/dx_{F})_{\textrm{two-comp}}}
  {(dn/dx_{F})_{\textrm{inclus}}} ,
\end{equation}
as presented in Fig.~\ref{fig:fover_shape}.

\begin{figure}[b]
  \centering
  \epsfig{file=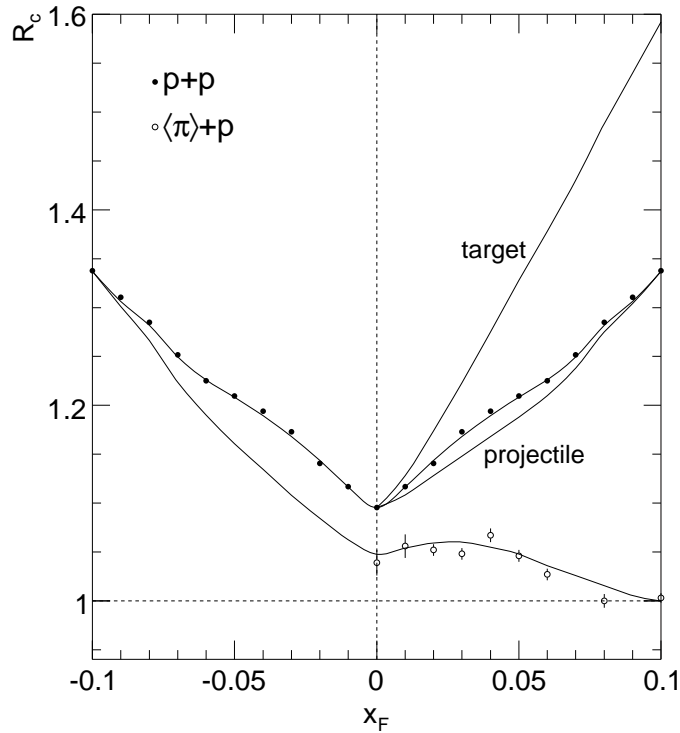,width=9cm}
  \caption{$\pi^+$/$\pi^-$ ratio in p+p and $\langle \pi \rangle$+p with
           corresponding predictions for the proton target and proton 
           projectile components }
  \label{fig:meas_rat}
\end{figure}

In this Fig. two slightly different shapes are indicated. They
correspond to the two pion charges. This shape difference is
necessitated by the structure in the $\pi^+$/$\pi^-$ ratio around 
$x_F$~=~0.04 observed in Fig.~\ref{fig:pipr_rat}. In fact there is 
no a priori knowledge of the behaviour of the two charge states in the 
feed-over region: they may and indeed do differ from each other. This 
difference is quantified in Fig.~\ref{fig:fover_shape}b and c with respect 
to the charge-average distribution. It corresponds to a decrease followed by 
a steep increase of the $\pi^+$/$\pi^-$ ratio as one moves from the target 
hemisphere into the projectile region. This is shown in Fig.~\ref{fig:meas_rat} 
which gives the corresponding prediction for the measured $\pi^+$/$\pi^-$ 
ratio both for p+p and $\langle \pi \rangle$+p collisions.

It is interesting to note that the measurement for both reactions 
is precisely described, yielding a simultaneous explanation for
the slight shoulder and the structure around $x_F$~=~0.04 observed
in p+p and $\langle \pi \rangle$+p collisions, respectively. 
This corresponds to a long-range charge correlation
within the target and projectile contributions with the respective
incoming hadron, which is for the first time extracted here.
This correlation will present an interesting challenge for 
microscopic production models, in particular for those based on
string fragmentation ideas, as the memory of the original parton
charge tends to be quickly lost along the string \cite{bib:field}. 

%
%
\subsection{Comparison to the feed-over in net baryon production}
\vspace{3mm}
\label{sec:baryon}

\begin{figure}[h]
  \centering
  \epsfig{file=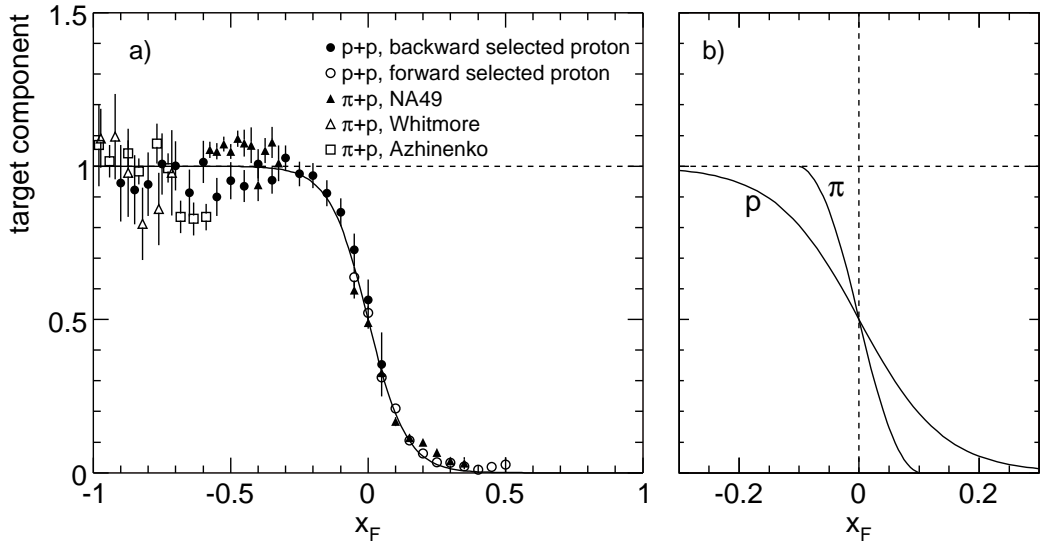,width=14cm}
  \caption{a) Target component for net protons measured with different methods
           and b) target components of $\langle \pi \rangle$ and net protons }
  \label{fig:pipr_fover}
\end{figure}

The NA49 data from p+p and $\langle \pi \rangle$+p interactions have been 
previously used to extract the two-component nature of net baryon production.
The method used is different from the one applied for the pion
production as net baryon number conservation can be invoked as a
powerful constraint. The argumentation gives consistent results
using pion induced collisions and the forward-backward baryon
correlation in p+p events \cite{bib:dezso}. The extracted shape of the feed-over
distribution is directly and independently measured by three methods
in a completely model-independent way. The result is shown in 
Fig.~\ref{fig:pipr_fover} in comparison to the pion feed-over obtained above.

There is clearly an important difference between net protons and
pions concerning the range of the feed-over. This indicates a
strong mass dependence. In connection with the yield increase
with transverse momentum, Sect.~\ref{sec:pt_dist}, it will be argued that there
is also a $p_T$ dependence of the feed-over. Both the mass and the
$p_T$ dependence will be shown in Sect.~10 to follow from resonance decay.
 
%
%
\section{Projectile and target components in the p+C interaction}
\vspace{3mm}
\label{sec:pc_two}

%
%
\subsection{Pion yields}
\vspace{3mm}
\label{sec:pc_two_yields}

The two-component picture of elementary hadronic collisions developed
in the preceding section can now be used to construct the superposition
of target and projectile components also for p+C interactions. The third
component given by intranuclear cascading will be shown in 
Sect.~\ref{sec:casc} below to not contribute in the studied $x_F$ range.
 
Several straight-forward assumptions have to be made in order to
quantify this superposition:

\begin{itemize} 
\item Both the target and the projectile components are assumed to follow
      the same feed-over mechanism as extracted for the elementary 
      collisions as far as relative shape and $x_F$ range are concerned, 
      as shown in Fig.~\ref{fig:fover_shape}.
\item In terms of pion yields the target contribution has however a weight
      $\langle \nu \rangle$ where $\nu$ is the number of projectile 
      collisions with the participating target nucleons. The underlying 
      assumption is here that subsequent projectile collisions lead to 
      the same density distribution per participating nucleon.
\item In addition and due to the isoscalar nature of the Carbon nucleus, 
      the density distribution for both $\pi^+$ and $\pi^-$ is built up as:

      \begin{equation}
	\left(\frac{dn}{dx_F}\right)_{\!\!\textrm{target}}(\pi^\pm) = 
	\frac{1}{2}\left(\frac{dn}{dx_F}\right)^
	     {\!\!\textrm{pp}}_{\!\!\textrm{target}}(\pi^+) 
	+ \frac{1}{2}\left(\frac{dn}{dx_F}\right)^
	{\!\!\textrm{pp}}_{\!\!\textrm{target}}(\pi^-) 
      \end{equation}
      which results as necessary in a $\pi^+$/$\pi^-$ ratio of unity over 
      the full $x_F$ range.
\item The projectile component on the other hand carries the full 
      imprint of $\langle \nu \rangle$ collisions and is as such not 
      predictable. Its model-independent quantification and extraction 
      is the main aim of this section. In a first instance, it is assumed 
      to be identical to the elementary p+p collision,

      \begin{equation} 
        \left(\frac{dn}{dx_F}\right)_{\!\!\textrm{proj}}(\pi^\pm) =
        \left(\frac{dn}{dx_F}\right)^{\!\!\textrm{pp}}_{\!\!\textrm{proj}}(\pi^\pm)
        \label{eq:proj}
      \end{equation}
      with the aim to extract its real shape by inspecting the deviations
      of the data from this simple parametrization.
\end{itemize}
 
With these assumptions, a prediction for the expected pion density
distribution in p+C collisions can be obtained as follows:
 
\begin{equation}
  \left(\frac{dn}{dx_F}\right)_{\!\!\textrm{pred}}(x_F) = 
  \langle \nu \rangle\left(\frac{dn}{dx_F}\right)_{\!\!\textrm{target}}(x_F) 
  + \left(\frac{dn}{dx_F}\right)^{\!\!\textrm{pp}}_{\!\!\textrm{proj}}(x_F)
\end{equation}
 
For ease of comparison with the data, see Sect.~\ref{sec:comp_pp}, the 
prediction is referred to the p+p interaction by forming the ratio
 
\begin{equation}
  R_{\textrm{pred}}(x_F) = \frac{(dn/dx_F)_{\textrm{pred}}(x_F)}
  {dn/dx_F^{\textrm{pp}}(x_F)}
\end{equation}
 
\begin{figure}[h]
  \centering
  \epsfig{file=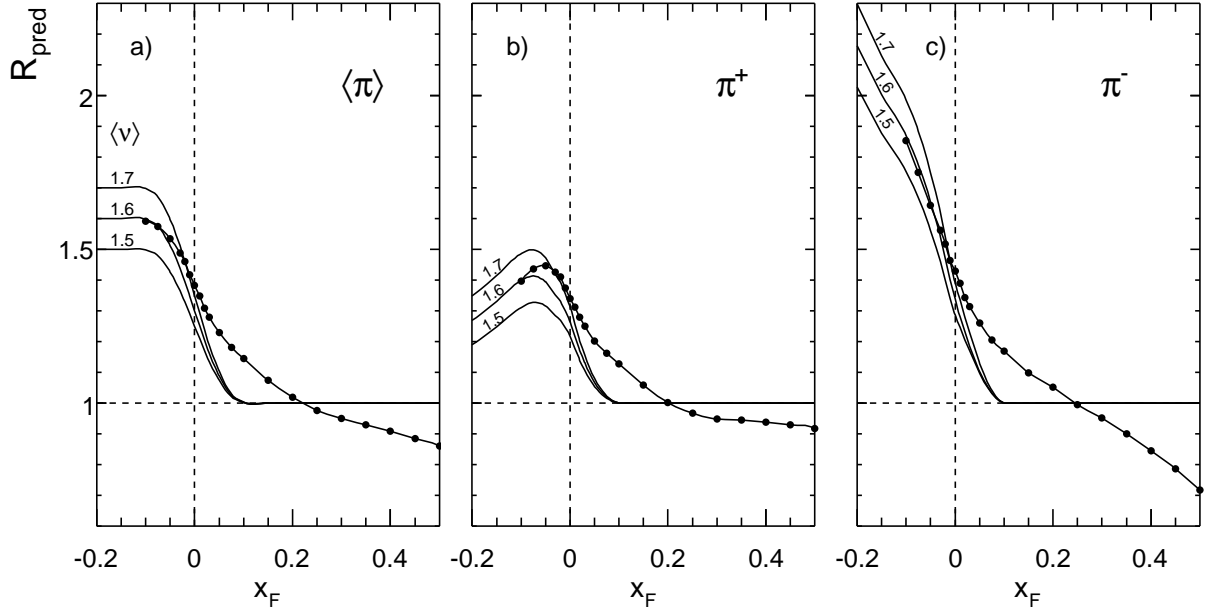,width=16cm}
  \caption{$R_{\textrm{pred}}$ for different $\langle \nu \rangle$ 
           compared to the measurement (circles) for a) $\langle \pi \rangle$,
	   b) $\pi^+$, and c) $\pi^-$; the data points, 
	   $R(x_F) = (dn/dx_F)^{\textrm{pC}}/(dn/dx_F)^{\textrm{pp}}$ 
	   are also given}
  \label{fig:pc_two}
\end{figure}

This ratio is plotted in Fig.~\ref{fig:pc_two} for three different values 
of $\langle \nu \rangle$ and compared to the measurements as a function of $x_F$.

Several points are noteworthy in this comparison:

\begin{itemize} 
\item A fair description of the data is obtained in the backward hemisphere
      at $x_F$ towards -0.1 for the value $\langle \nu \rangle$~=~1.6. 
      The sensitivity of the comparison to $\langle \nu \rangle$ is clearly 
      evident from the Fig.~\ref{fig:pc_two}.
\item This holds simultaneously for the average pion yield and for the
      $\pi^+$ and $\pi^-$ yields separately, both in size and in local slope
\item This is a strong indication for the validity of the two-component
      picture in this region where the target contribution alone should
      determine the total yields for $x_F$~=~-0.1 and below.
\item Moving from this $x_F$ value towards the forward region, the 
      measurements exceed $R_{\textrm{pred}}$ up to 
      $x_F \sim$~0.2 in a characteristic fashion.
\item For $x_F >$~0.2 the data fall below the predicted values 
      $R_{\textrm{pred}}$.
\end{itemize}
 
\begin{figure}[h]
  \centering
  \epsfig{file=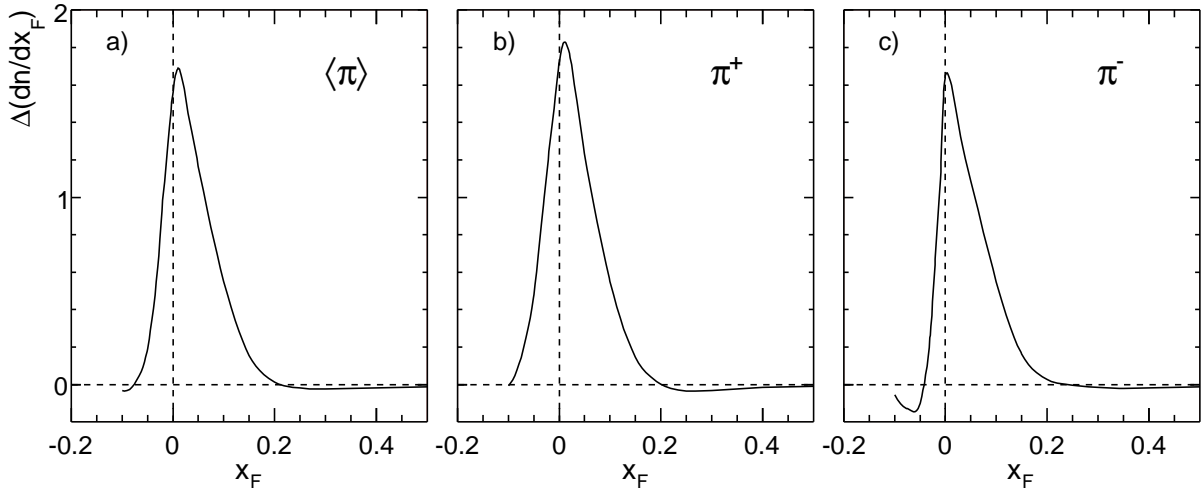,width=16cm}
  \caption{Difference between measured ratio and predicted yield for 
           a) $\langle \pi \rangle$, b) $\pi^+$, and c) $\pi^-$ for 
	   $\langle \nu \rangle$ = 1.6}
  \label{fig:pred_dev}
\end{figure}

The deviation from $R_{\textrm{pred}}$ can be translated into 
absolute particle densities as shown in Fig.~\ref{fig:pred_dev} as a function 
of $x_F$.
  
\begin{figure}[h]
  \centering
  \epsfig{file=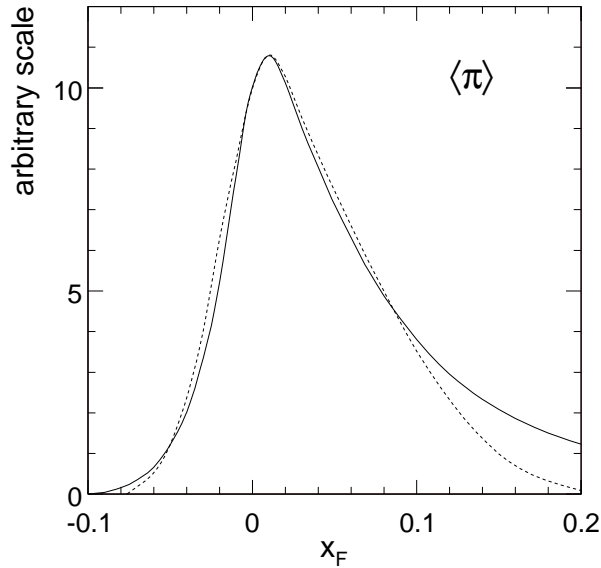,width=8cm}
  \caption{Shape comparison of the $x_F$ distribution of $\Delta (dn/dx_F)$ in p+C 
           (dashed line) and the projectile component $(dn/dx_F)_\textrm{proj}$ 
           in p+p collisions (full line) }
\label{fig:shape}
\end{figure}
  
The obtained density difference follows the shape of the projectile 
contributions in p+p collisions up to $x_F \sim$~0.1. This is shown in 
Fig.~\ref{fig:shape} where the difference is superimposed to the 
projectile charge density distribution for $\langle \pi \rangle$, 
see Fig.~\ref{fig:pp_2comp}. This is a further strong indication that 
the two-component superposition picture produces indeed a valid description 
of the data. The region of reduced particle density at $x_F >$~0.2 contributes 
as expected only a very small negative contribution to the absolute increase 
of pion density. Integrating the density differences of 
Fig.~\ref{fig:pred_dev} it emerges that the projectile 
pion yields are enhanced by about 10\% with respect to p+p collisions 
\cite{bib:pp_paper}. This observation carries interesting
consequences for the interpretation of the corresponding increase of
pion production observed in nucleus-nucleus interactions \cite{bib:dezso1}.
 
%
%
\subsection{Charge ratios}
\vspace{3mm}
 
The investigation of the $\pi^+$/$\pi^-$ ratio offers an additional and very
sensitive tool for the scrutiny of the target-projectile superposition
mechanism. As already shown in Sect.~\ref{sec:comp_pp}, Fig.~\ref{fig:c2p_rat}, 
the overall behaviour of the charge ratio complies with the expectation 
from a two-component picture in the sense that unity is approached in the 
region of prevailing target contribution, $x_F <$~-0.1, and the 
$\pi^+$/$\pi^-$ ratio observed in p+p collisions is approached in the 
projectile fragmentation region, $x_F >$~0.1. There are, however, detailed 
deviations from this first-order expectation. The approach to unity at 
negative $x_F$ is not smooth but resembles the structure observed in 
$\langle \pi \rangle$+p collisions at $x_F >$~0, i.e. in the corresponding
isoscalar projectile fragmentation region of the elementary hadronic reaction.
The $\pi^+$/$\pi^-$ ratio of p+p collisions is not quite reached in the forward
region below $x_F$~=~0.3. These deviations correspond to a modification of
the projectile charge ratio in multiple collision processes both in
the forward and backward regions. A first indication for the backward
hemisphere is contained in Fig.~\ref{fig:pred_dev} where the density 
difference for $\pi^+$ reaches further out towards negative $x_F$ 
than the one for $\pi^-$. This resembles the behaviour already found for the 
elementary collisions, Sect.~\ref{sec:pp_two}. 
                                                                               
\begin{figure}[h]
  \centering
  \epsfig{file=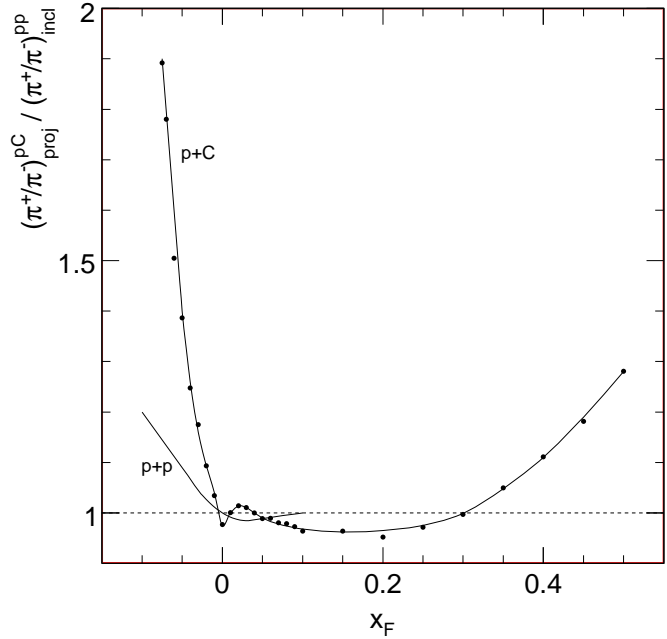,width=9cm}
  \caption{Charge ratio of the projectile component in p+C divided by the 
           inclusive ratio in p+p collisions }
  \label{fig:dev_rat}
\end{figure}

Using the quantitative superposition scheme developed in the preceding
Sect.~\ref{sec:pc_two_yields} for $\langle \nu \rangle$~=~1.6, which 
explicitly contains the $\pi^+$/$\pi^-$ ratio of unity for the target 
fragmentation part, the deviation of the projectile charge ratio from 
the one measured in inclusive p+p collisions may be extracted. It is 
shown in Fig.~\ref{fig:dev_rat} as the ratio 
$(\pi^+/\pi^-)^{\textrm{pC}}_{\textrm{proj}}/
(\pi^+/\pi^-)^{\textrm{pp}}_{\textrm{incl}}$
  
Apparently there is again an increase of the charge ratio in the
projectile feed-over region. This increase is stronger than the
one observed in elementary collisions, Fig.~\ref{fig:meas_rat}, 
also shown for comparison. It is followed by a decrease of about 4\% 
between $x_F \sim$~0.1  and $x_F \sim$~0.25 and again by an increase at 
$x_F$ beyond 0.3. This represents an interesting feature of the 
multiple collision process in proton-nucleus collisions in the sense that 
the projectile charge is correlated to both extremes of the fragmentation 
region. Due to the small contribution in absolute yield from these areas,
the decrease of the charge ratio in the intermediate region
is imposed by total charge conservation in view of the increase
of projectile multiplicity.
 
%
%
\section{Nuclear aspects}
\vspace{3mm}
\label{sec:aspects}

Up to this point the discussion was performed without recourse to any
details concerning the manifestation of the Carbon nucleus. Only its
isospin symmetry and the general possibility of multiple projectile
collisions were invoked. It is now necessary to come to a more
quantitative assessment of a number of nuclear parameters which are
essential for the further extension of the argumentation. The decisive
quantity for most of the nuclear parameters is the nuclear density
distribution, $\rho(r)$, which determines the rms radius and the
probability of single projectile collision, $P(1)$. The second quantity 
of principle interest is the projectile-nucleon total inelastic cross section,
or rather its evolution with subsequent collisions. Only by
making very distinct assumptions about this quantity further
important parameters like the probability distribution of multiple
collisions, $P(\nu)$ and its mean value $\langle \nu \rangle$ , 
already introduced above, may be obtained. The quantitative results on 
several parameters discussed below come from a detailed Monte Carlo 
calculation \cite{bib:andrzej} using the density profile as an input.
 
%
%
\subsection{Nuclear density distribution and percentage of single collisions}
\vspace{3mm}
 
\begin{figure}[h]
  \centering
  \epsfig{file=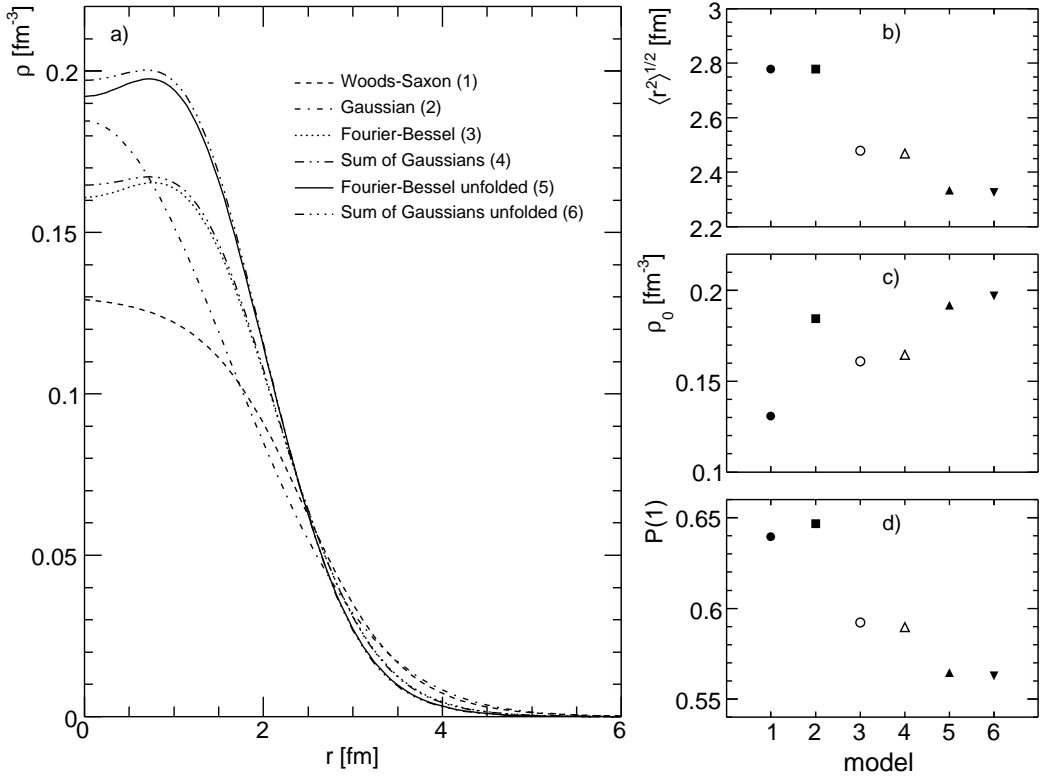,width=14cm}
  \caption{a) Nuclear density profile parametrizations; b) rms radius; 
           c) nuclear density at $r$~=~0; d) probability for single collision }
\label{fig:nuc_dens}
\end{figure}

The principle experimental information available for the Carbon nucleus
comes from electron scattering experiments \cite{bib:el_scat} which 
measure the nuclear charge distribution. This is not identical with the 
density profile $\rho(r)$ as the spatial distribution of charge in the 
proton has to be unfolded. An additional problem resides in the possible 
asymmetry between proton and neutron distributions. Altogether this 
leads to a substantial uncertainty about the density profile.
Fig.~\ref{fig:nuc_dens} gives $\rho(r)$ for six different parametrizations:

\begin{enumerate} 
\item Saxon-Woods \cite{bib:book}.
\item Single Gaussian \cite{bib:andrzej}.
\item Fourier-Bessel \cite{bib:fricke}.
\item Sum of Gaussians \cite{bib:sick}.
\item Fourier-Bessel unfolded for proton charge distribution \cite{bib:andrzej}. 
\item Sum of Gaussians unfolded for proton charge distribution \cite{bib:andrzej}.
\end{enumerate} 
 
\begin{figure}[h]
  \centering
  \epsfig{file=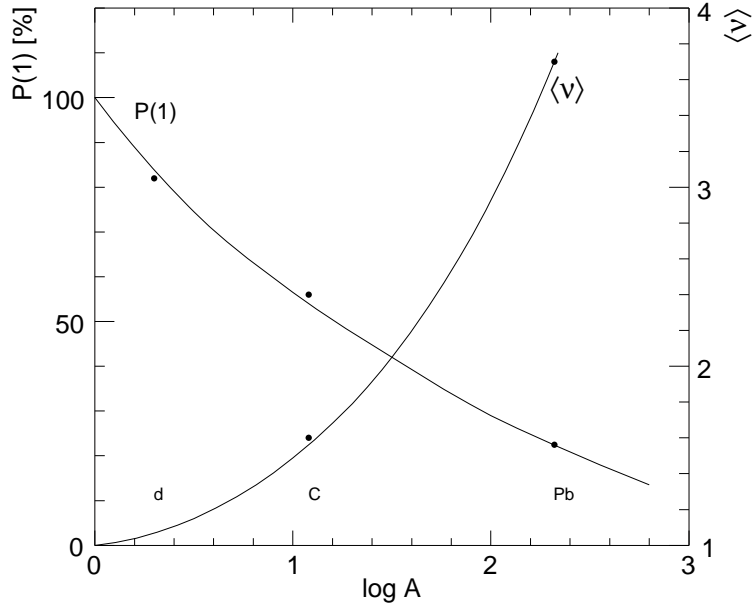,width=10cm}
  \caption{ Probability for single collision and mean number of collisions as
            a function of atomic number}
\label{fig:nu_prob}
\end{figure}

In consequence the rms radius, the nuclear density at the origin
and the probability of single collisions (which depends on the
elementary proton+nucleon inelastic cross section) show substantial
spreads according to the parametrization chosen. This is quantified
in the Figs.~\ref{fig:nuc_dens}b-d for the rms radius, the central 
nucleon density $\rho$($r$=0), and the probability of single collision
$P(1)$. Excluding the first two options which are known to disagree 
with the experimental evidence, one may conclude on an rms radius 
between 2.3 and 2.5 fm and on a fraction of single collisions between 
0.56 and 0.60. The latter quantity recalls a basic weakness of 
minimum bias proton-nucleus experiments given by the fact that a large 
fraction of events corresponds to ``trivial'' proton-nucleon collisions. 
This fraction descends only slowly with increasing atomic number A 
\cite{bib:andrzej}, as shown in Fig.~\ref{fig:nu_prob}.
  
The contribution of single collisions can in principle be taken out
of the experimental results since the percentage $P(1)$ is rather
precisely determined and the corresponding target and projectile
components can be safely predicted from elementary collisions.
This will be demonstrated in the subsequent sections of this paper.
An experimentally more efficient way is given by triggering on
the presence of ``grey'' protons \cite{bib:hh1} already at the level 
of data taking which has the additional advantage of permitting the 
control of centrality. Unfortunately, this method could not be applied 
to the present data due to external constraints on the time available 
for data taking \cite{bib:pc_paper}.
 
%
%
\subsection{The probability distribution of multiple hits and its relation to
            the total inelastic cross section}
\vspace{3mm}
 
\begin{figure}[t]
  \centering
  \epsfig{file=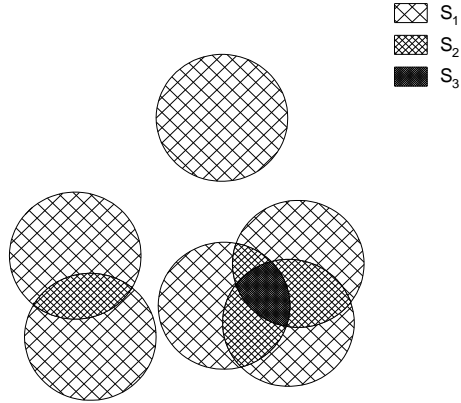,width=7cm}
  \caption{Schematic view of the superposition of nucleon-nucleon 
           cross sections in a nuclear environment }
  \label{fig:sketch}
\end{figure}

Any further specification of nuclear parameters in relation to the
interaction of a hadronic projectile with the nucleus has to rely on
assumptions on the elementary hadron-nucleon cross section for
multiple collisions. The problem may be visualized in a simple
geometrical way by looking at a random superposition of A circular
disks, each with a surface S of $\sigma^{\textrm{pp}}$~=~31.8 mb, 
as shown in Fig.~\ref{fig:sketch}.

The surface area covered by one layer of disks is denoted by S1.
S2 is the area covered by two layers, S3 by three layers etc.
Scanning the plane with a random distribution of projectiles,
the mean hit number of layers will be

\begin{equation} 
  \langle \nu \rangle = \frac{S_1 + 2S_2 + 3S_3 + \ldots + AS_A}
                             {S_1 + S_2 + S_3 + \ldots + S_A}.
\end{equation}
 
The total area covered by the disks is $S_1 + S_2 + S_3 + \ldots + S_A$. 
It represents the total inelastic cross section $\sigma^{\textrm{pA}}$. 
The total surface of the disks, which is $A\sigma^{\textrm{pp}}$, 
is given by

\begin{equation} 
  A\sigma^{\textrm{pp}} = S_1 + 2S_2 + 3S_3 + \ldots + AS_A.
\end{equation}
 
The mean number of collisions is therefore defined by the formula 
\cite{bib:bialas}
 
\begin{equation} 
  \langle \nu \rangle = \frac{A\sigma^{\textrm{pp}}}
	  {\sigma^{\textrm{pA}}}.
\end{equation}
 
\begin{figure}[b]
  \centering
  \epsfig{file=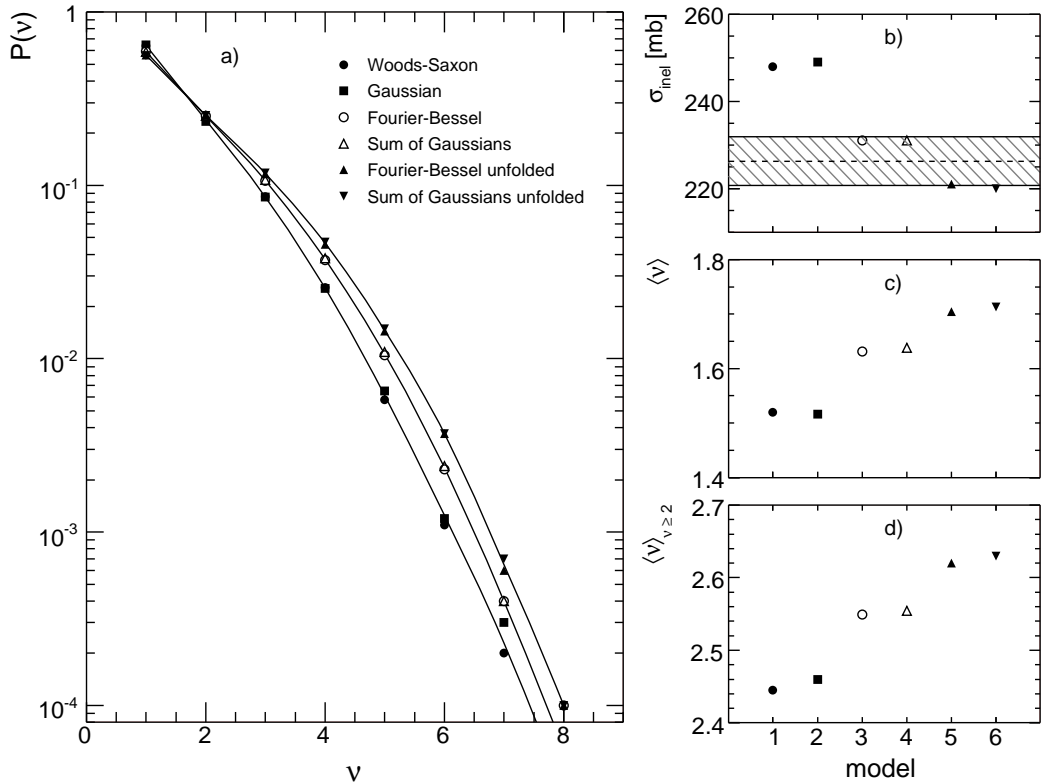,width=14cm}
  \caption{a) Probability of $\nu$ collisions for different nuclear density 
           parametrizations; b) inelastic cross section. The dashed line 
	   indicates the NA49 result with the error margin (shaded area); 
	   c) mean number of collisions; 
	   d) mean number of collisions for $\nu \geq$~2 }
  \label{fig:ncoll_prob}
\end{figure}
  
This formula is clearly only valid if the disk
surface describing the inelastic projectile-nucleon cross
section is the same in each of the successive layers corresponding
to the multiple projectile collisions. Any experimental
verification of the mean number of collisions provides therefore
an important constraint on the evolution of the cross section
with $\nu$.
 
Always under the above assumption, the distribution $P(\nu)$ of the probability
to produce $\nu$ collisions in one event, has been determined, using the
different nuclear density profiles discussed above, via a  Monte Carlo
calculation \cite{bib:andrzej}. The result is shown in 
Fig.~\ref{fig:ncoll_prob} which also gives the integrated quantities 
$\sigma_{\textrm{inel}}^{\textrm{pA}}$, 
$\langle \nu \rangle$ and the mean number of collisions excluding single 
hits, $\langle \nu  \rangle_{\nu \geq 2}$.
  
In principle the measured value of the inelastic cross section at
226.3~mb~$\pm$~2\% (Fig.~\ref{fig:ncoll_prob}b) gives a constraint on the 
nuclear profile and favours the charge-unfolded distributions (5) and (6). 
This would then correspond to a predicted mean number of collisions of
about 1.68. The value of 1.6 extracted from the pion data in the preceding 
section is indeed close but significantly below this prediction 
as can be seen from the strong dependence on $\langle \nu \rangle$ of
the density ratios shown in Fig.~\ref{fig:pc_two}.

%
%
\section{Intranuclear cascading}
\vspace{3mm}
\label{sec:casc} 

\begin{figure}[b]
  \centering
  \epsfig{file=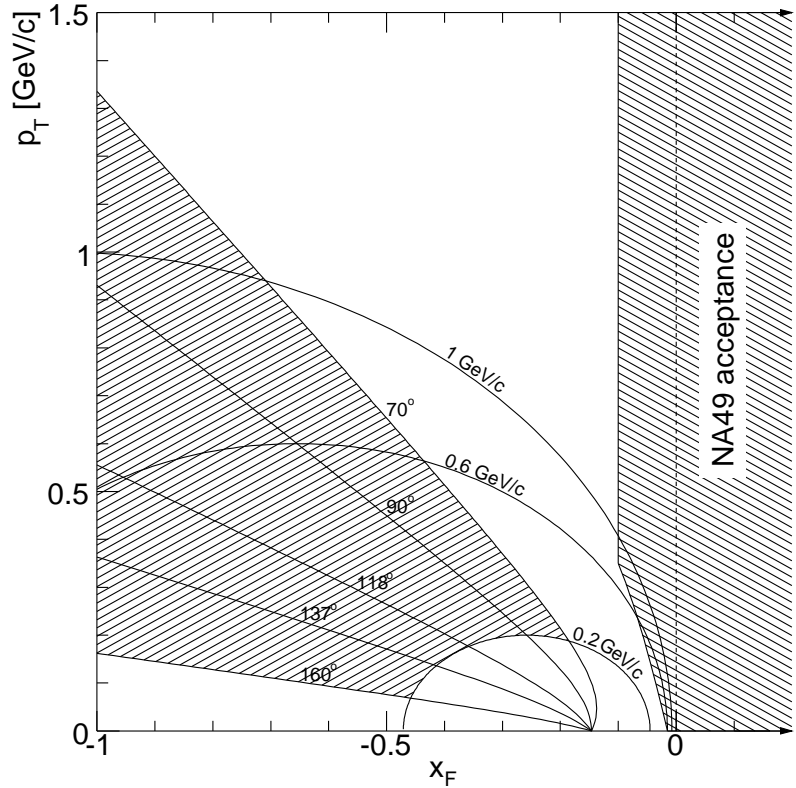,width=11cm}
  \caption{Kinematic situation in the backward region. With lines are indicated
           the constant total momentum $p_{\textrm{lab}}$ (0.2, 0.6 
           and 1 GeV/c) and constant polar angle $\Theta_{\textrm{lab}}$
	   (70$^o$, 90$^o$, 118$^o$, 137$^o$, 160$^o$). The shaded areas represent
           the coverage of \cite{bib:niki} (far backward region) and 
	   NA49 (central and forward region). }
\label{fig:casc_kin}
\end{figure}

In the preceding discussion only the principle components of
target and projectile fragmentation have been treated. The
third contribution to the pion yields due to the propagation
and interaction of the participant nucleons and of their
fragmentation products inside the nucleus, will now be
investigated. The proper Lorentz frame for the description of
this intranuclear cascading process is the rest system of the
target as the cascading products will have typically small
momenta in this system. The kinematic situation is indicated
in Fig.~\ref{fig:casc_kin}. Here lines at constant total momentum 
$p_{\textrm{lab}}$ and constant polar angles 
$\Theta_{\textrm{lab}}$ for pions in the target rest
frame are given in the $x_F$/$p_T$ plane. Pion production is centered
at low $p_{\textrm{tot}}$ at $x_F \sim$~-0.15, as expected 
from the mass ratio $m_\pi$/$m_p$. Large values of 
$p_{\textrm{lab}}$ are needed to approach, at forward 
production angles, the region close to $x_F$~=~0.
  
The NA49 acceptance is indicated as a hatched area in 
Fig.~\ref{fig:casc_kin}. It appears that this acceptance matches 
in its backward low $p_T$ part the line at 0.6 GeV/c total momentum. 
At this momentum the intranuclear pion production yield is already reduced 
and only small if any cascading contributions are to be expected from
this kinematic fact alone. This argument can however be quantified
by making use of the measurements of Nikiforov et al. \cite{bib:niki} 
at Fermilab with a beam momentum of 400 GeV/c. This important and
unique experiment covers the backward region in the second hatched
area indicated in Fig.~\ref{fig:casc_kin} by measuring identified pion 
yields at constant $\Theta_{\textrm{lab}}$ between 70 and 160 
degrees, leaving only a small uncovered zone with respect to the NA49 
acceptance. A set of 32 cross sections is available at $x_F >$~-1.0 
for the assessment of pion yields in p+C interactions.

\begin{figure}[b]
  \centering
  \epsfig{file=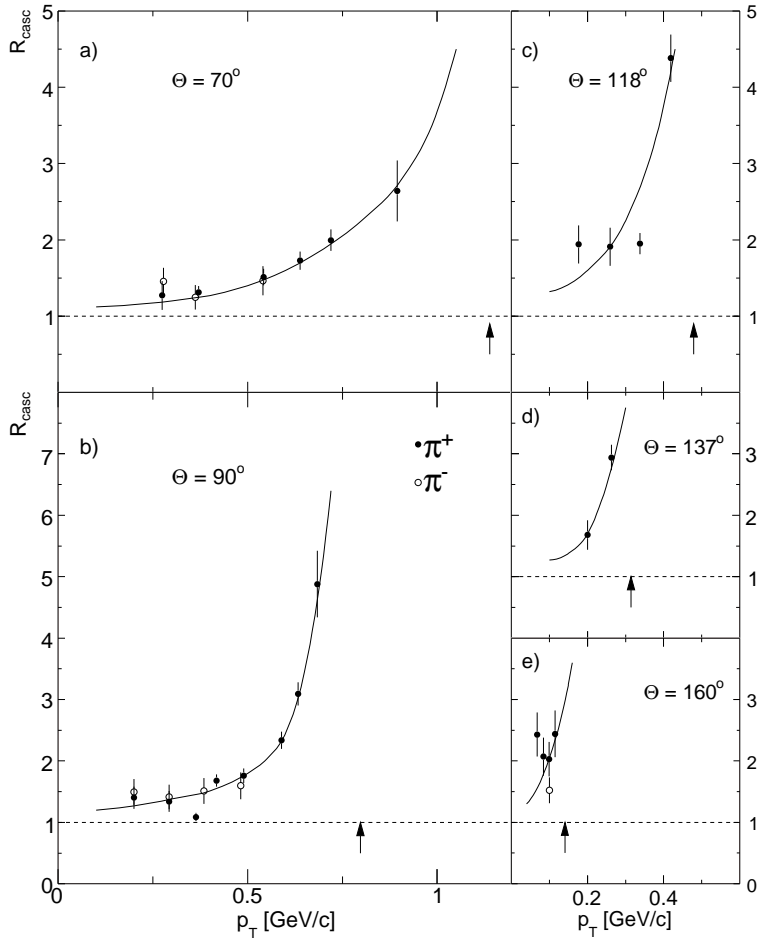,width=10cm}
  \caption{$R_{\textrm{casc}}$ for different fixed lab angles
           $\Theta_{\textrm{lab}}$ measured by \cite{bib:niki}.
           The arrows indicate the $p_T$ value at which $x_F$ of -0.85 is 
	   reached }
\label{fig:casc}
\end{figure}
 
As the experimental layout of the experiment \cite{bib:niki} does not 
allow a discrimination of off-vertex tracks it must however be expected
that the complete feed-down from weak decays of strange particles
is contained in the data at these low total momenta. A feed-down
correction has therefore been calculated and applied following
the methodology developed for the NA49 data \cite{bib:pc_paper}.
    
In order to put the Nikiforov et al. data \cite{bib:niki} into perspective 
and to allow for a quantitative comparison, the given cross sections are
related to the isospin averaged prediction for the target
fragmentation developed and justified in the preceding sections.
This is possible since no contribution from projectile hadronization
can be expected in the covered area. The ratio
 
\begin{equation}
  R_{\textrm{casc}} = \frac{(Ed^3\sigma/dp^3)_{\textrm{Nikiforov}}}
       {(Ed^3\sigma/dp^3)_{\textrm{pC}}}
\end{equation}   
is plotted in Fig.~\ref{fig:casc} for the 5 measured angles as a 
function of $p_T$.
  
$R_{\textrm{casc}}$ increases strongly with $p_T$ for 
all angles. This is due to the corresponding increase of $|x_F|$ along 
the lines of constant $\Theta_{\textrm{lab}}$ together with 
the increasing dominance of cascading products in the far backward region. 
In fact cascading products are found at $x_F <$~-1.0 which is not accessible 
for target fragmentation. $R_{\textrm{casc}}$ therefore tends
to diverge towards this limit. The arrows given in Fig.~\ref{fig:casc} 
indicate the $p_T$ value at which $x_F$ of -0.85 is reached.
 
\begin{figure}[h]
  \centering
  \epsfig{file=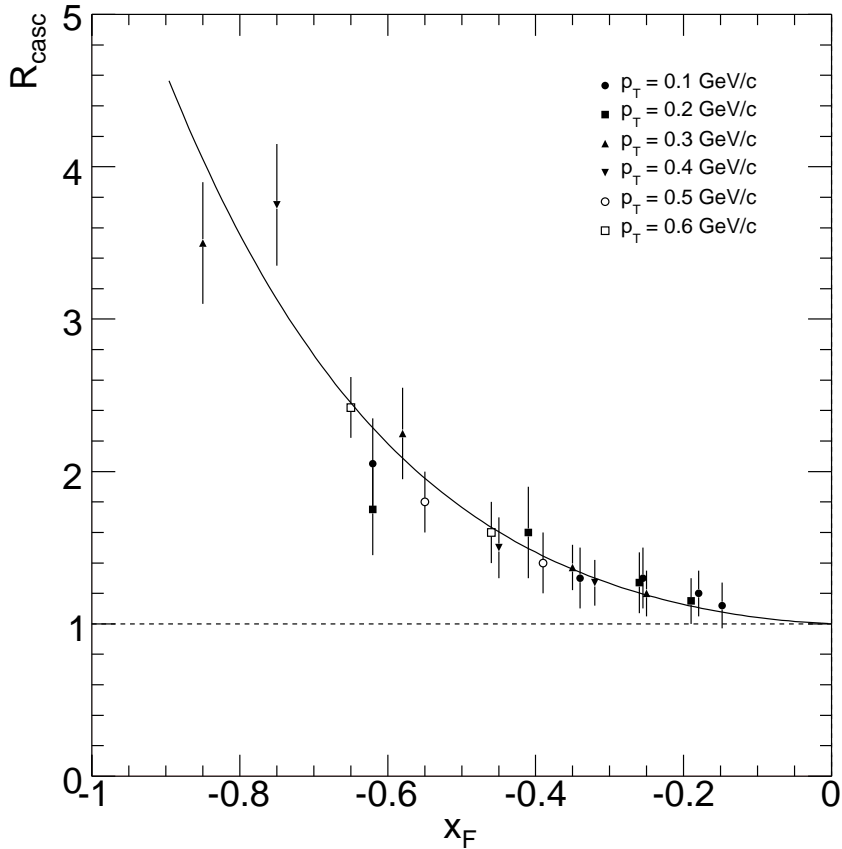,width=12cm}
  \caption{$R_{\textrm{casc}}$ obtained from \cite{bib:niki} as a function 
           of $x_F$}
  \label{fig:casc_xf}
\end{figure}

The dependence on $p_T$ at fixed $\Theta_{\textrm{lab}}$ can be 
transformed into a dependence on $x_F$ at fixed $p_T$ by using eye-ball 
fits through the data points of Fig.~\ref{fig:casc} for interpolation. 
The result is shown in Fig.~\ref{fig:casc_xf}.
  
Within the sizeable error bars of the data points and the interpolation
used, a universal curve appears which clearly shows convergence towards
$R_{\textrm{casc}}$~=~1 at $x_F$~=~0. In fact the constraint 
of $R_{\textrm{casc}}$ to 1 at $x_F$~=~0 is justified by 
the fact that relatively large values of $p_{\textrm{lab}}$ are 
requested to approach $x_F$~=~0, see Fig.~\ref{fig:casc_kin}. As the ratio 
$R_{\textrm{casc}}$ gives directly the excess of particle 
yield due to cascading with respect to target fragmentation, it may be 
argued that the cascading contribution is only on a few percent level at 
$x_F$~=~-0.1, that is, at the NA49 acceptance limit. The fact that this 
acceptance limit applies to $p_T$~=~0.4~GeV/c or 
$p_{\textrm{lab}}$~=~0.6~GeV/c further reduces the possible 
contribution to the NA49 data. The extrapolation into the unmeasured 
region used in the preceding sections, which is oriented towards target 
fragmentation alone, is therefore justified. It should be mentioned 
here that in the region $x_F <$~-0.1 and $p_T <$~0.2, effects concerning
the $\pi^+$/$\pi^-$ ratio must be present due to the final-state Coulomb
interaction with the spectator protons in the nucleus \cite{bib:coulomb}. 
These effects are however expected to be small for the light Carbon nucleus.
 
\begin{figure}[h]
  \centering
  \epsfig{file=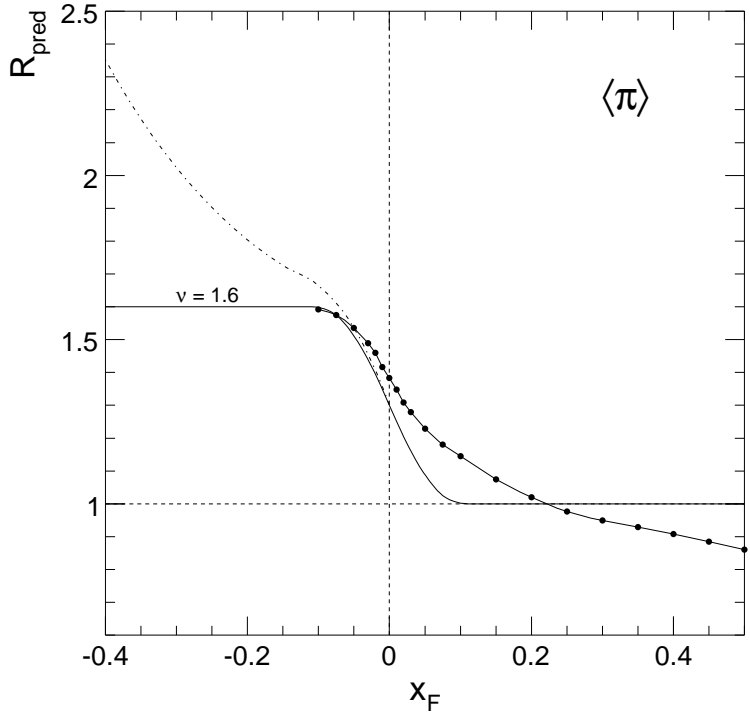,width=10cm}
  \caption{$R_{\textrm{pred}}$ as a function of $x_F$, giving the data 
           points together with the target contribution at 
           $\langle \nu \rangle$~=~1.6 (full line) and adding the 
	   intranuclear cascading part (broken line) }
  \label{fig:casc_pred}
\end{figure}

In Sect.~\ref{sec:pc_two} above a quantitative prediction from target and 
projectile fragmentation alone has been given extending into the far backward
region $x_F <$~-0.1. This prediction must now be completed taking account
of the cascading contribution as quantified in Fig.~\ref{fig:casc_xf} above. 
This correction has been performed for the pion yield ratio 
$R_{\textrm{pred}}$ shown in Fig.~\ref{fig:pc_two} for the 
charge averaged pion yields where it developed a plateau in this region. 
The complete prediction is presented in Fig.~\ref{fig:casc_pred}.
  
In this plot, all three principle contributions to the total pion
yield, i.e. projectile and target fragmentation, and nuclear cascading,
are clearly discernible and quantified. The resulting overall ratio
to the elementary collision exhibits a smooth increase from the
far forward to the far backward region in $x_F$. This is in agreement
with the pseudo-rapidity $\eta$ ratio

\begin{equation} 
  R(\eta) = \frac{(dn/d\eta)^{\textrm{pA}}(\eta)}{(dn/d\eta)^{\textrm{pN}}(\eta)}
\end{equation}
observed in emulsion work \cite{bib:abduz} at 200 and 800 GeV/c beam 
momentum. Also here a quasi-linear increase of $R(\eta)$ is observed, 
from values below 1 at forward rapidity to values in excess of 2 in
the far backward hemisphere.
 
%
%
\section{Contribution from multiple collisions}
\vspace{3mm}
\label{sec:mult_coll}

As discussed in Sect.~\ref{sec:aspects} and due to the important tail 
of the nuclear density profile of the Carbon nucleus, a large fraction
of all events correspond to a single proton-nucleon collision in
a minimum bias p+C experiment. This probability $P(1)$ is located
between 56 and 60\% corresponding to a reasonable choice of nuclear
density distributions (see Sect.~\ref{sec:aspects} and 
Fig.~\ref{fig:nuc_dens}). Although these single collisions can be 
accompanied by some amount of nuclear cascading due to the single 
recoil nucleon, this component will not influence the measurements 
in the NA49 acceptance region (see Sect.~\ref{sec:casc}). Given $P(1)$, 
it is therefore possible to extract from the data the content of multiple 
collisions with $\nu >$~1 which will then rather correspond to a mean value 
$\langle \nu \rangle \sim$~2.5 (Fig.~\ref{fig:ncoll_prob}).
 
\begin{figure}[h]
  \centering
  \epsfig{file=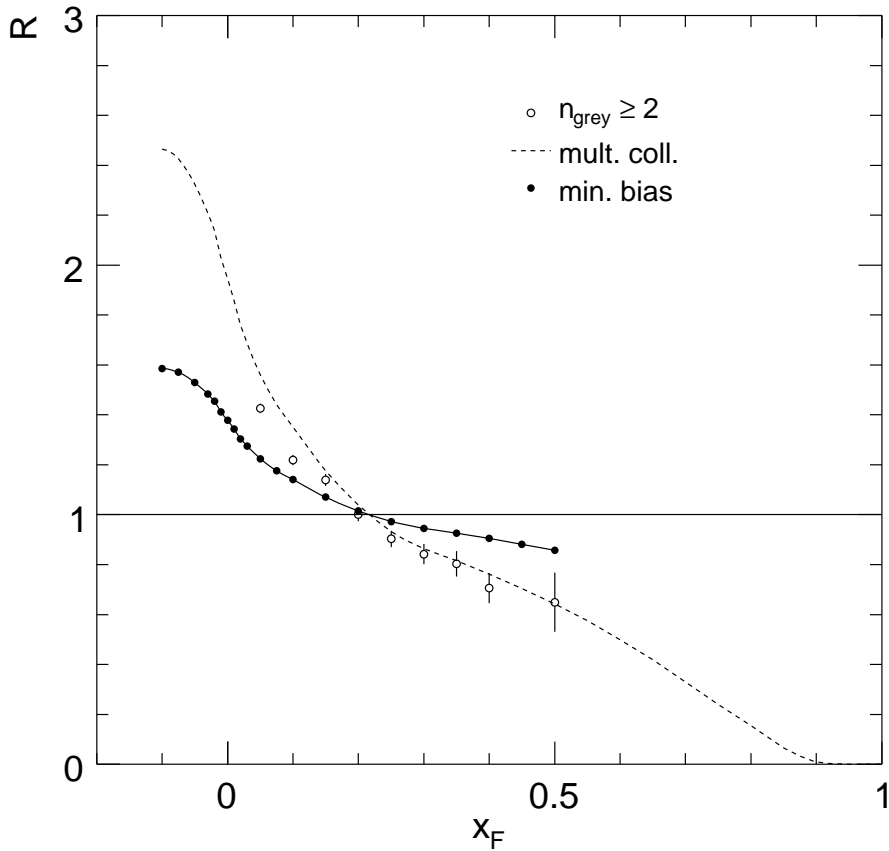,width=12cm}
  \caption{Prediction for $R_{\textrm{mult}}(x_F)$ compared
           with measurement for minimum bias and $n_{grey} \geq$~2  }
  \label{fig:mult_rat}
\end{figure}

In a first example the extraction is applied to the charge averaged
density ratio $R(x_F)$ as defined in Eq.\ref{eq:yield_rat} above. In this 
case the multiple collision component is defined as

\begin{equation} 
  R_{\textrm{mult}}(x_F) = 
  \frac{(dn/dx_F)^{\textrm{pC}}_{\textrm{mult}}(x_F)}{(dn/dx_F)^{\textrm{pC}}(x_F)} 
   = \frac{R(x_F) - P(1)}{1-P(1)}     
\end{equation}
and is presented in Fig.~\ref{fig:mult_rat} as a function of $x_F$.
  
In comparison to the minimum bias measurement, 
$R_{\textrm{mult}}(x_F)$ steepens up and reaches the 
values 2.5 at $x_F$~=~-0.1 and 0.65 at $x_F$~=~0.5. It is
interesting to regard the limit $x_F \rightarrow$~1. Even if the 
suppression of pion density in the far forward region of p+A collisions 
did not as yet find a satisfying explanation, most of the possibilities 
will finally empty the large $x_F$ region of pions from multiple 
collisions. This is true for projectile energy loss where the scale 
of $x_F$ will change, as well as for baryon transfer processes (``stopping''),
should baryonic resonance decay be the source of forward pions.
In these cases $R_{\textrm{mult}}(x_F)$ will approach 
zero and $R(x_F \rightarrow 1)$ will reach $P(1)$. This can finally 
provide a direct measurement of this probability. Fig.~\ref{fig:mult_rat} 
contains in this sense a tentative extrapolation of 
$R_{\textrm{mult}}(x_F)$ towards zero to illustrate the 
situation which will be further constrained by the upcoming publication 
of proton data in p+C interactions and of comparable data from p+Pb 
collisions.
 
A sharp test of this extraction procedure is given by the data sample
with centrality constraint also available from NA49 \cite{bib:pc_paper}. 
Here the additional measurement of slow (``grey'') protons in the lab 
momentum range below 1.2~GeV/c provides an event sample where single 
collisions should be rather effectively suppressed. The data points 
shown in Fig.~\ref{fig:mult_rat} come from events with at least two 
grey protons detected. As the pion densities for this relatively small 
subsample have been determined in \cite{bib:pc_paper}, the ratio
 
\begin{equation} 
  R_{\textrm{grey}}(x_F) = \frac{(dn/dx_F)^{\textrm{pC}}_{\textrm{grey}}(x_F)}
  {(dn/dx_F)^{\textrm{pp}}(x_F)}    
\end{equation}

can be obtained. Evidently this measurement approaches the predicted
behaviour for multiple collisions rather closely and proves both the
effectiveness of this way of experimental centrality control 
\cite{bib:hh1,bib:nim} and the reliability of the prediction of 
$P(1)$ from nuclear parameters.
 
\begin{figure}[h]
  \centering
  \epsfig{file=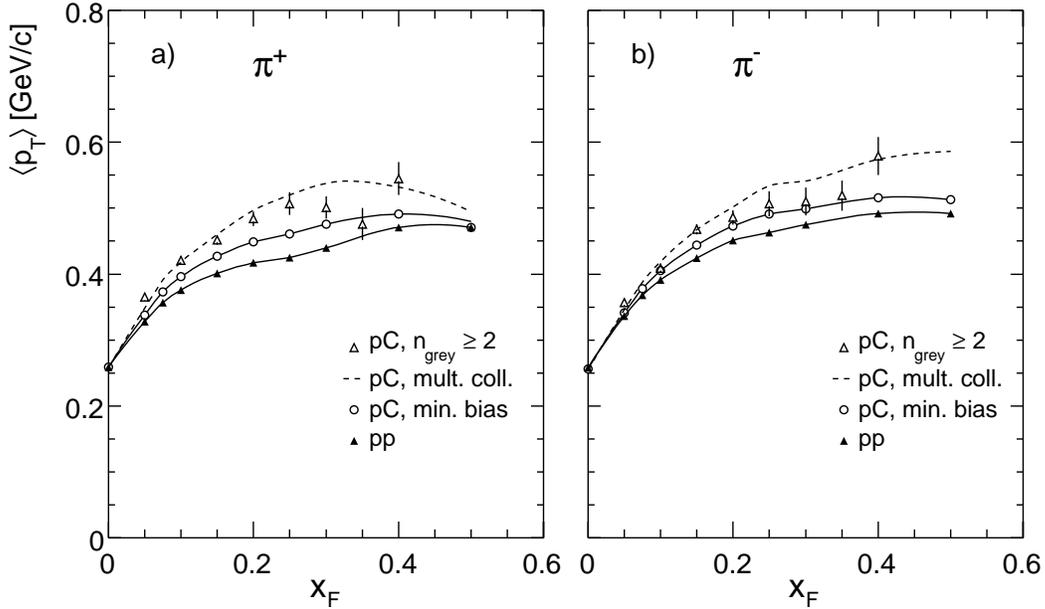,width=14cm}
  \caption{Prediction for $\langle p_T \rangle$ in multiple collisions compared
           with measurement in p+p, p+C minimum bias, and p+C with selected number
           of grey protons $n_{grey} \geq$~2 }
  \label{fig:mult_meanpt}
\end{figure}

As a second example for the extraction of improved information on
multiple collisions the evolution of the mean transverse momentum
with multiple collisions will be discussed. The situation is shown
in Fig.~\ref{fig:mult_meanpt} where $\langle p_T \rangle$ is shown for 
p+p collisions \cite{bib:pp_paper}, minimum bias p+C interactions 
\cite{bib:pc_paper} and for the event sample with at least two
grey protons discussed above.
  
Evidently the increase of $\langle p_T \rangle$ from elementary to minimum 
bias nuclear reactions is accentuated imposing centrality control. The multiple
collision part can also be extracted from the single collision probability
$P(1)$ via the formula
 
\begin{equation} 
  \langle p_T \rangle^{\textrm{mult}} = 
  \frac{\langle p_T \rangle^{\textrm{pC}} -
    P(1) r \langle p_T \rangle^{\textrm{pp}}}{1 - P(1) r},          
\end{equation}
where r~=~$(dn/dx_F)^{\textrm{pp}}/(dn/dx_F)^{\textrm{pC}}$~=~$1/R(x_F)$.
                                                                               
The result is shown as the upper line in Fig.~\ref{fig:mult_meanpt}. Again 
both the measurement with grey proton constraint and the prediction from 
nuclear parameters are coming close.
                                                                               
A third example of this kind will be discussed in connection with the
increase of pion yield with transverse momentum in Sect.~\ref{sec:pt_dist} below.

%
%
\part{$\mathbf{p_T}$ dependence}
\vspace{3mm}
\label{part:two}
Extending the discussion from $p_T$ integrated information to $p_T$ dependent
quantities opens up a new and more complex subsurface of the
multidimensional phase space as it is projected into the double inclusive
cross section

\begin{equation} 
  f(x_F,p_T) = E(x_F,p_T) \cdot \frac{d^3\sigma}{dp^3} (x_F,p_T)
\end{equation}
 
Indeed already the superficial inspection of this quantity in its
$x_F$ and $p_T$ dependence for p+p \cite{bib:pp_paper} and 
p+C \cite {bib:pc_paper}interactions reveals significant local structure 
which has a different $x_F$ and $p_T$ dependence for the two reactions. 
In the following sections only a restricted number of phenomena can be 
touched upon, with the aim at bringing out as clearly as possible some 
important features which are specific to p+A collisions and therefore 
also relevant for the subsequent work on p+Pb and Pb+Pb interactions. 
The discussion will be conducted again in relation to the precise
elementary reference given by the NA49 data on p+p collisions 
\cite{bib:pp_paper}.
  
%
%
\section{Cross section ratio with respect to elementary collisions as a function of $\mathbf{p_T}$}
\vspace{3mm}
\label{sec:pt_dist}   

%
%
\subsection{Definition of the cross section ratio}
\vspace{3mm}

The presence of the Carbon nucleus and of the corresponding two-component
hadronization mechanism discussed in the preceding sections implies that the
straight-forward cross section ratio
 
\begin{equation}
  R(x_F,p_T) = \frac{f^{\textrm{pC}}(x_F,p_T)}
                    {f^{\textrm{pp}}(x_F,p_T)} 
\end{equation} 

is probably not the best choice for describing the full $x_F$ scale. There
are in fact two important and well quantified overall phenomena which
should and can be taken out in the comparison:

\begin{itemize} 
\item The target pile-up corresponding to the 1.6 participant nucleons
      which introduces an $x_F$-dependent overall upward shift of $R$
      in the region $x_F <$~0.1 as elaborated in Sect.~\ref{sec:pc_two}.
\item The isoscalar nature of the Carbon nucleus which averages both pion
      charges in the target contribution, as opposed to the projectile
      fragmentation.
\end{itemize}
 
\begin{figure}[h]
  \centering
  \epsfig{file=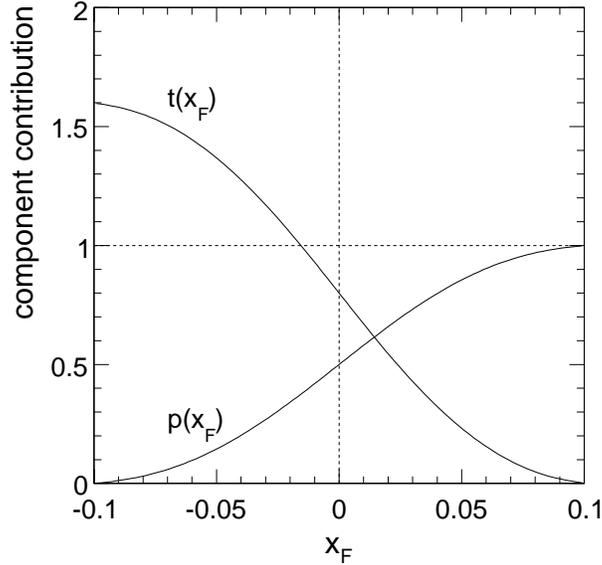,width=8cm}
  \caption{Target $t(x_F) $and projectile $p(x_F)$ contributions to the pion yield
           in the two-component picture, with $t(x_F)$ = 1.6$p(-x_F)$ }
  \label{fig:targ_proj}
\end{figure}

In order to characterize the projectile-connected $p_T$ dependencies in an
optimal way, a definition of the cross section ratio that takes direct
reference to the two-component superposition, has been chosen:

\begin{equation} 
  R_{p_T}(x_F,p_T) = \frac{f^{\textrm{pC}}(x_F,p_T)}
                          {f_{\textrm{two-comp}}(x_F,p_T)}
\label{eq:rpt}
\end{equation}
where $f_{\textrm{two-comp}}(x_F,p_T)$ is defined as:

\begin{equation}  
  f_{\textrm{two-comp}}(x_F,p_T) = 
  t(x_F)\left(\frac{1}{2}f_{\pi^+}^{\textrm{pp}}(x_F,p_T) +
  \frac{1}{2}f_{\pi^-}^{\textrm{pp}}(x_F,p_T)\right) + 
  p(x_F)f^{\textrm{pp}}(x_F,p_T),
\end{equation}
with $f^{\textrm{pp}}(x_F,p_T)$ denoting the double differential 
pion cross section in p+p interactions. The functions $t(x_F)$ and $p(x_F)$ 
describe the relative contribution of the target and projectile fragmentation 
to the overall pion yield in the two-component picture, Sects.~\ref{sec:pp_two} 
and \ref{sec:pc_two} above, as recalled in Fig.~\ref{fig:targ_proj}.
  
\begin{figure}[t]
  \centering
  \epsfig{file=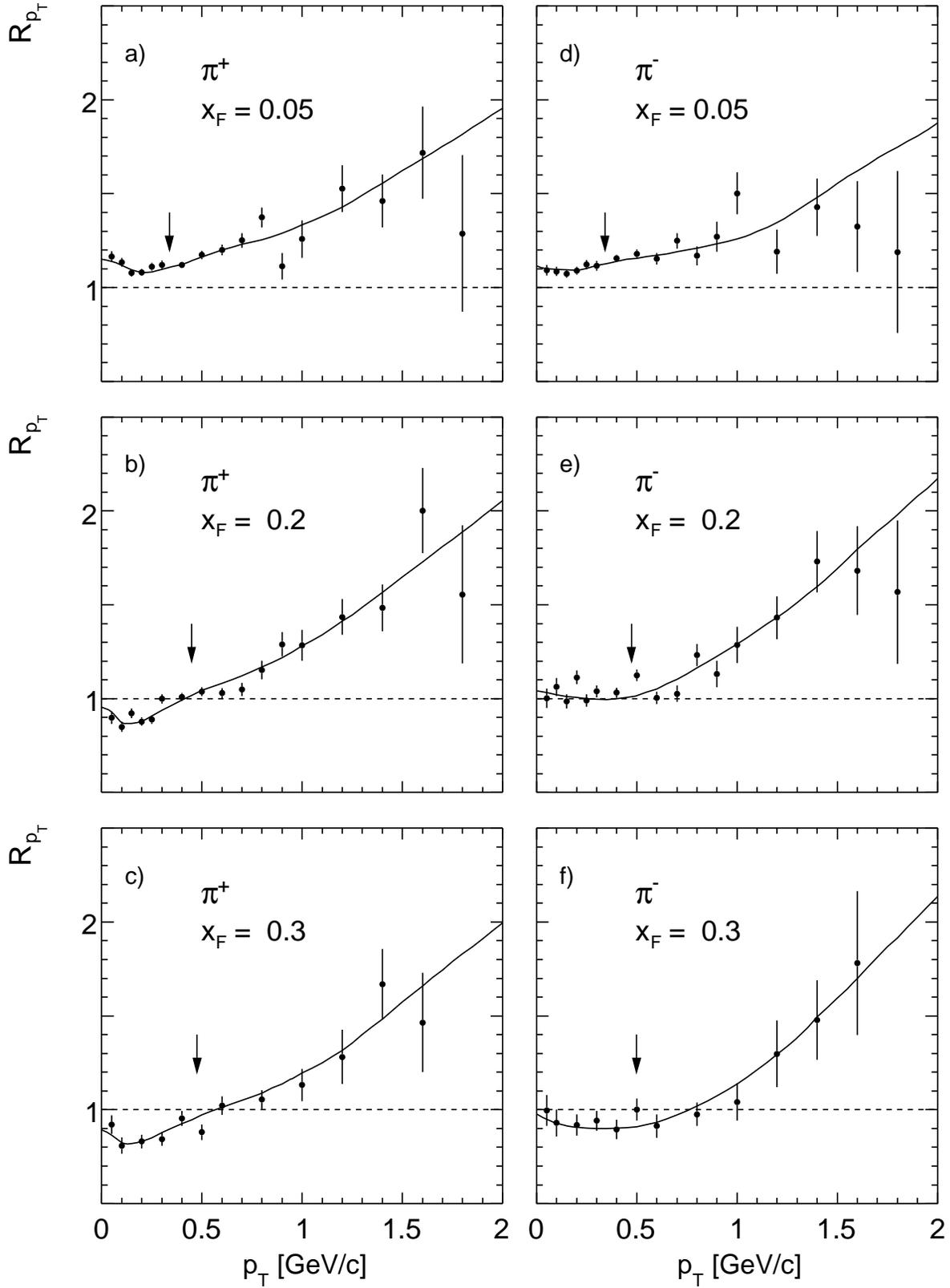,width=16cm}
  \caption{Ratio $R_{p_T}$ as a function of $p_T$ at three $x_F$ values for
           a),b) and c) $\pi^+$, and d),e) and f) $\pi^-$. The arrows indicate
           $\langle p_T \rangle$ for given $x_F$ }
  \label{fig:pt_c2p_ex}
\end{figure}

Evidently, $t(x_F)$ ensures that the target contribution amounts to
$\langle \nu \rangle$ times the average pion yield at $x_F <$~-0.1 and 
vanishes at $x_F$~=~0.1. On the other hand $p(x_F)$ determines the total 
yield at $x_F >$~0.1 and vanishes towards $x_F$~=~-0.1. The ratio 
$R_{p_T}$ therefore contains the salient features of the 
superposition of elementary components in the p+C interaction and 
addresses specifically the modifications due to the projectile 
fragmentation. This is shown in Fig.~\ref{fig:pt_c2p_ex} for three values 
of $x_F$ and for both $\pi^+$ and $\pi^-$.
  
Here the full lines represent $R_{p_T}$ using the $x_F$/$p_T$ interpolation 
developed for the p+p \cite{bib:pp_paper} and p+C \cite{bib:pc_paper} 
interactions. As expected, complex $p_T$ and $x_F$ dependencies emerge. 
The $p_T$ range around the mean values $\langle p_T \rangle$,
indicated by arrows in the plots, governs the $p_T$ integrated yields.
It reproduces the features of this quantity in relation to the elementary
collisions as elaborated above, namely an increase below $x_F$~=~0.2
followed by a zero-crossing and a decrease in the far forward region.

\begin{figure}[t]
  \centering
  \epsfig{file=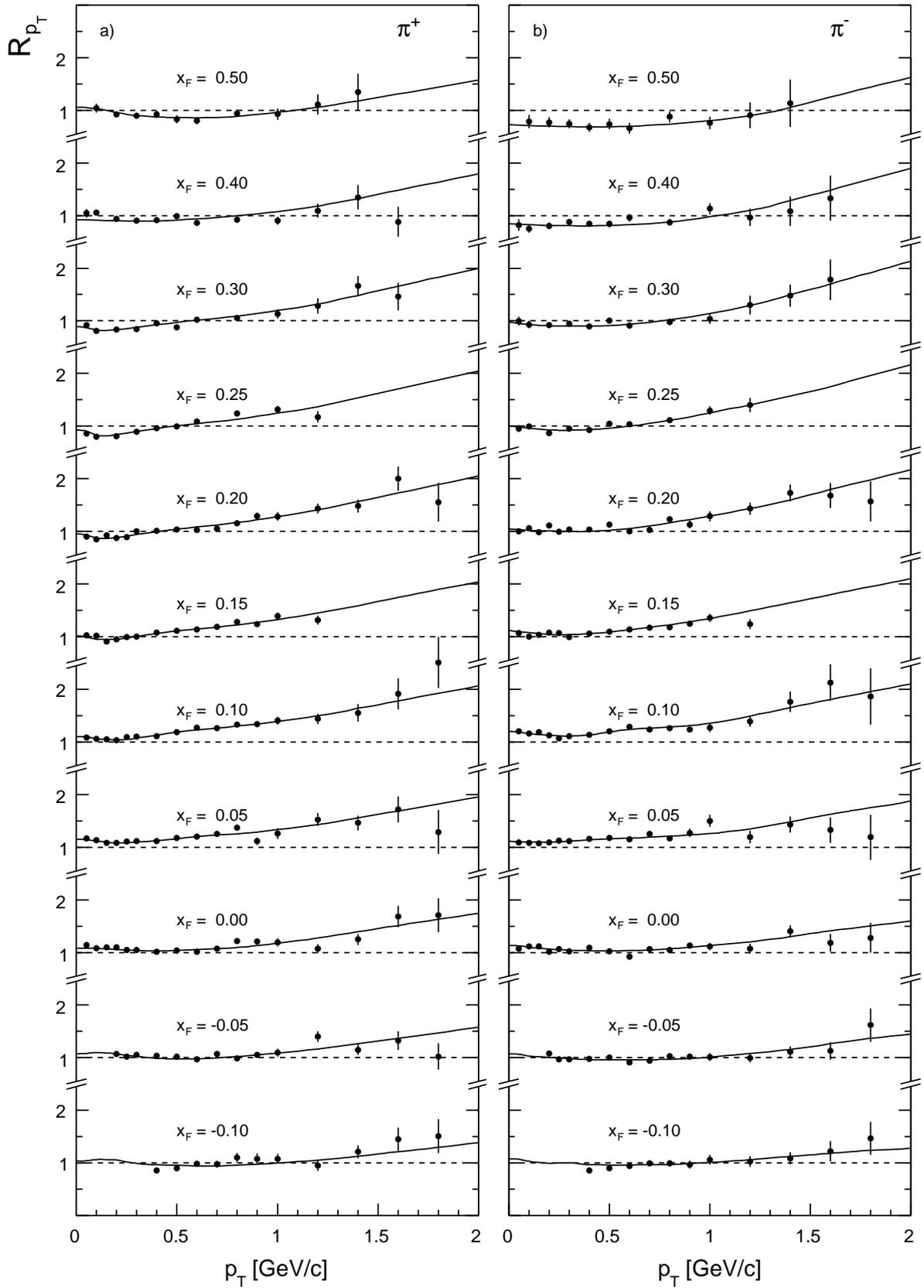,width=16cm}
  \caption{Ratio $R_{p_T}$ as a function of $p_T$ at fixed $x_F$ for a) $\pi^+$
           and b) $\pi^-$ }
  \label{fig:rpt}
\end{figure}

The relative $p_T$ dependence of $R_{p_T}$ in the vicinity of $\langle p_T \rangle$
shows a rather complicated evolution. As the fine structure of 
$f^{\textrm{pp}}(x_F,p_T)$ and $f^{\textrm{pC}}(x_F,p_T)$
has been connected to resonance production and decay \cite{bib:pp_paper}, 
this means that the principle source of pion production changes its composition 
in p+A interactions both as a function of $x_F$ and of $p_T$, and in a 
different way for the two pion charges. Here the attention should be drawn 
specifically to the behaviour at low $p_T$ governed by the cascading decay 
of low mass resonances.
  
A second region of interest is the evolution towards high values of $p_T$.
A significant increase of $R_{p_T}$ with respect to $\langle p_T \rangle$ is 
evident over the full $x_F$ scale as shown in Fig.~\ref{fig:rpt} which 
provides an overview of $R_{p_T}(x_F,p_T)$ covering the full range of the 
measurements.
  
%
%
\subsection{Definition of the $\mathbf{p_T}$ enhancement}
\vspace{3mm}
 
In order to assess the increase of the pion yields as a function of
transverse momentum, the measured enhancement factor $E^m_{p_T}$ is defined as

\begin{equation} 
  E^m_{p_T}(p_T) = \frac{R_{p_T}(p_T)}{R_{p_T}(\langle p_T \rangle)}
\end{equation}
 
\begin{figure}[h]
  \centering
  \epsfig{file=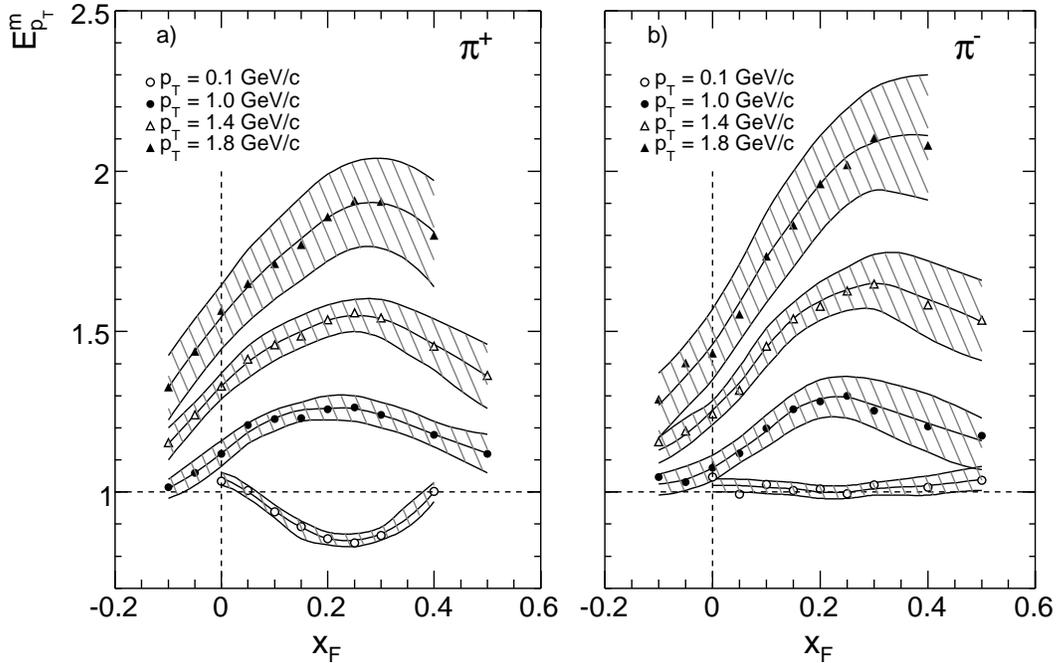,width=14cm}
  \caption{The $p_T$ enhancement as a function of $x_F$ at several fixed $p_T$ values.
           The shaded regions mark the error margins connected to the interpolation
	   scheme \cite{bib:pp_paper,bib:pc_paper} }
  \label{fig:ept}
\end{figure}

This definition relates the behaviour at any $p_T$ to the mean transverse
momentum (see Sect.~\ref{sec:mult_coll}). By referring to an integrated 
quantity which is representative for the $p_T$ integrated observables 
discussed in Part~\ref{part:one}, the detailed structures which are 
present at low $p_T$ are averaged out in the reference. The $x_F$ dependence 
of $E^m_{p_T}$ is shown in Fig.~\ref{fig:ept} for four values of $p_T$ extending 
up to the experimental limit at 1.8~GeV/c, and using the interpolation shown 
in Fig.~\ref{fig:rpt} for the determination of  $E^m_{p_T}$.

The shaded regions in Fig.~\ref{fig:ept} correspond to the estimation of the
statistical error margins connected with the interpolation procedure.
The enhancement factor increases, from values close to unity but
significantly different for $\pi^+$ and $\pi^-$ at low $p_T$, up to about 2
at $p_T$~=~1.8 GeV/c. The increase shows  a strong $x_F$ dependence, rising
from the acceptance limit at $x_F$~=~-0.1 through $x_F$~=~0 towards the 
forward direction. At $x_F >$~0.2, 0.3, a decrease of $E^m_{p_T}$ is visible.
 
These effects will be discussed in more detail for the high $p_T$ limit
in the following section.

%
%
\section{High $\mathbf{p_T}$ phenomena}
\vspace{3mm}
\label{sec:high_pt}  
 
In connection with the discussion of $p_T$ dependencies, in particular for
proton-nucleus interactions, the term ``high $p_T$'' needs some specification.
 
In a first instance, it seems to be a generally accepted notion that
hadron production in the region of $p_T \sim$~2 GeV/c is already governed
by hard parton-parton scattering. This expectation was first developed
in connection with ISR data \cite{bib:negra} and is still heavily used in the
interpretation of very recent RHIC results \cite{bib:gaard}. The notion ``high'' 
therefore tends to be synonymous with the applicability of perturbative QCD, 
although there is no indication of any serious experimental proof for this 
assumption.
 
In a second instance, the early discovery of an ``anomalous'' $p_T$ enhancement
in proton-nucleus collisions, the so-called Cronin effect \cite{bib:cronin}, 
in a corresponding region of transverse momentum, seems to point again into 
the same direction. In fact the only theoretical attempts at an understanding 
of the Cronin effect are based on multiple partonic scattering 
\cite{bib:cron_par}. And again the present experimental information does not 
permit in the least to affirm this conjecture.
 
In the context of this paper the notion ``high'' $p_T$ describes the upper limit
of $p_T$ available to NA49 in the p+C interaction. The authors consider any
relation of this notion to perturbative effects as an unfounded speculation.
 
%
%
\subsection{Anomalous nuclear enhancement}
\vspace{3mm}
\label{sec:nuc_enh} 

\begin{figure}[b]
  \centering
  \epsfig{file=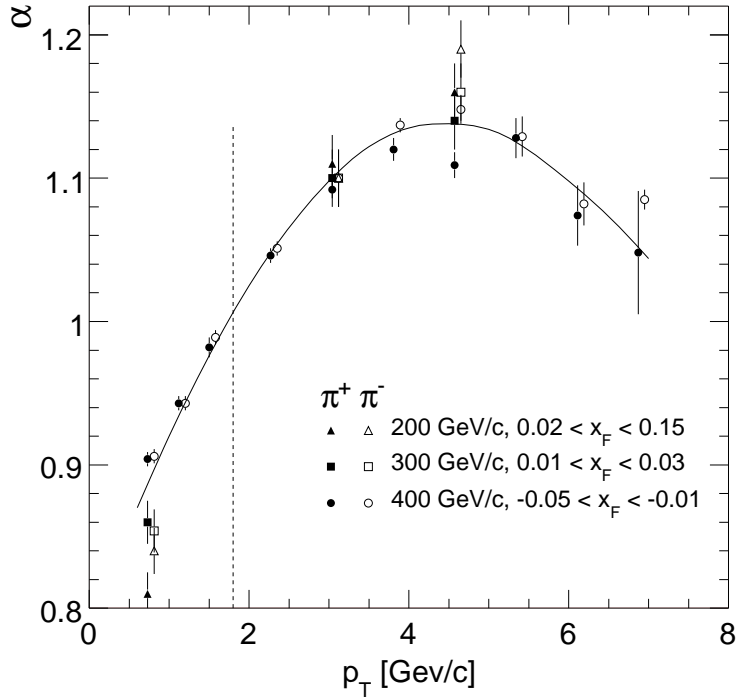,width=10cm}
  \caption{The exponent $\alpha$ as a function of $p_T$. The
           dashed line corresponds to $p_T$ limit of the NA49 data }
  \label{fig:alpha}
\end{figure}
  
The anomalous nuclear enhancement was first quantified via the A-dependence
of hadronic cross sections, parametrized as

\begin{equation} 
  \sigma^{\textrm{pA}} \sim A^{\alpha(p_T)}
\end{equation}
     
The dependence of the exponent $\alpha$ on transverse momentum \cite{bib:cronin} 
is recalled in Fig.~\ref{fig:alpha} for $\pi^+$ and $\pi^-$.
 
The exponent $\alpha(p_T)$ rises steeply from values between 0.8 and 0.9 at 
low $p_T$ to a maximum of about 1.15 at $p_T \sim$~4~GeV/c. Above this $p_T$ 
the exponent decreases again. The scarce measurements and the sizeable error 
bars at 200~GeV/c beam momentum, i.e. in the immediate vicinity of the NA49 
data set, do not exclude a sizeable $s$-dependence. It is also visible in 
Fig.~\ref{fig:alpha} that alpha is slightly larger for $\pi^-$ than for 
$\pi^+$. As regards the upper $p_T$ limit of the NA49 data at 1.8~GeV/c, 
Fig.~\ref{fig:alpha}, the corresponding $\alpha(p_T)$ is about half-way up 
from the low-$p_T$ value to the saturation limit. It happens to be very close 
to unity, a value that was believed to be characteristic of hard parton 
scattering \cite{bib:cronin}. The detailed study of the nuclear enhancement 
in this $p_T$ region gives therefore already valuable information concerning 
the understanding of the Cronin effect.
 
\begin{figure}[h]
  \centering
  \epsfig{file=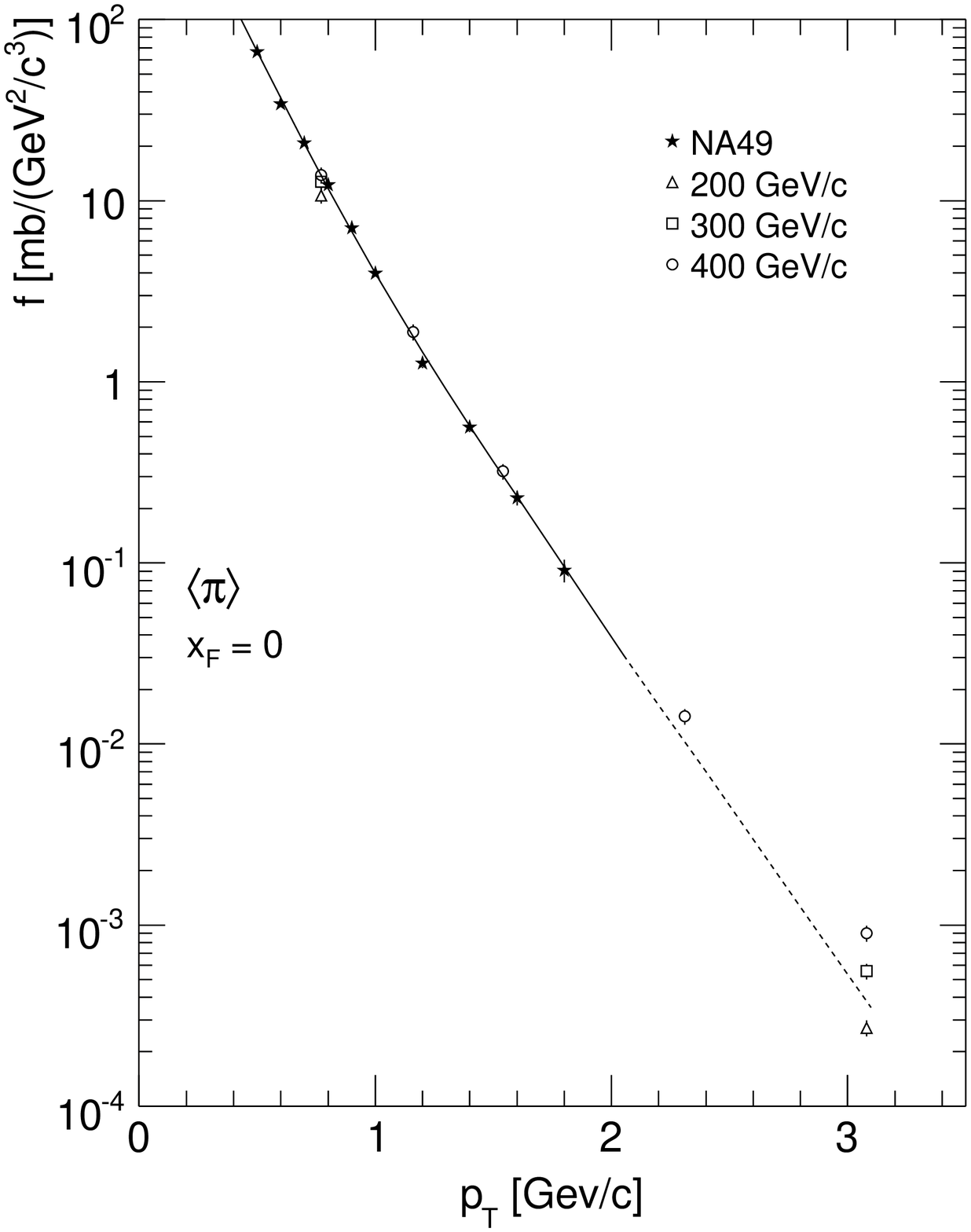,width=8cm}
  \caption{Comparison of the cross section for $\langle \pi \rangle$ as a 
           function of $p_T$ at $x_F$~=~0 measured by \cite{bib:cronin} and 
	   interpolated to Carbon with p+C results from NA49 }
\label{fig:cronin}
\end{figure}

The cross sections measured by \cite{bib:cronin} for Be, Ti and W nuclei 
may be interpolated to Carbon and compared directly to the NA49 data. This
comparison is shown in Fig.~\ref{fig:cronin} for $\langle \pi \rangle$.
  
Unfortunately there is only one overlapping data point at 200~GeV/c beam
momentum in the $p_T$ range of NA49, and the consistency of the few data
points at 400~GeV/c is not really expected due to the strong non-scaling
behaviour of the $s$-dependence in their $p_T$ range. In this context it 
must be recalled that the data of \cite{bib:cronin} have been obtained at 
a constant lab angle of 77~mrad which leads to an $s$-dependent spread of 
the corresponding $x_F$ values between -0.05 and +0.15, see 
Fig.~\ref{fig:alpha}. In view of the strong $x_F$ dependencies of both 
the target pile-up, Fig.~\ref{fig:casc_pred}, and the nuclear enhancement, 
Fig.~\ref{fig:cronin}, this leads to systematic effects which are 
non-negligible at the level of precision of the present study. Including 
however the 20\% normalization uncertainties quoted by \cite{bib:cronin} 
and the sizeable statistical errors of these measurements, the consistency 
of the data may still be considered as satisfactory. 

%
%
\subsection{Two basic contributions to the high $\mathbf{p_T}$ enhancement}
\vspace{3mm}
 
Turning back to the NA49 data, it may be claimed that they permit
the first systematic study of the $x_F$ dependence of the Cronin effect,
as the data of \cite{bib:cronin} only cover a small, $s$ and $p_T$ dependent
$x_F$ range. The new data at forward rapidity available from d+Au collisions 
at RHIC energy suffer from missing acceptance at low $p_T$. They will be 
commented on in a subsequent section of this paper.
 
Inspection of Fig.~\ref{fig:ept} shows that the central region of particle
production occupies by no means a special place, as the enhancement
rises smoothly through $x_F$~=~0 and has its maximum in the region around
$x_F$~=~0.3 which is far forward with respect to the longitudinal evolution
of pion yields. This behaviour indicates two important facts:

\begin{itemize}  
\item $E^m_{p_T}$ converges towards unity in the backward region, $x_F <$~-0.1.
\item $E^m_{p_T}$ tends as well towards unity in the far forward region, $x_F >$~0.4.
\end{itemize}
 
\begin{figure}[t]
  \centering
  \epsfig{file=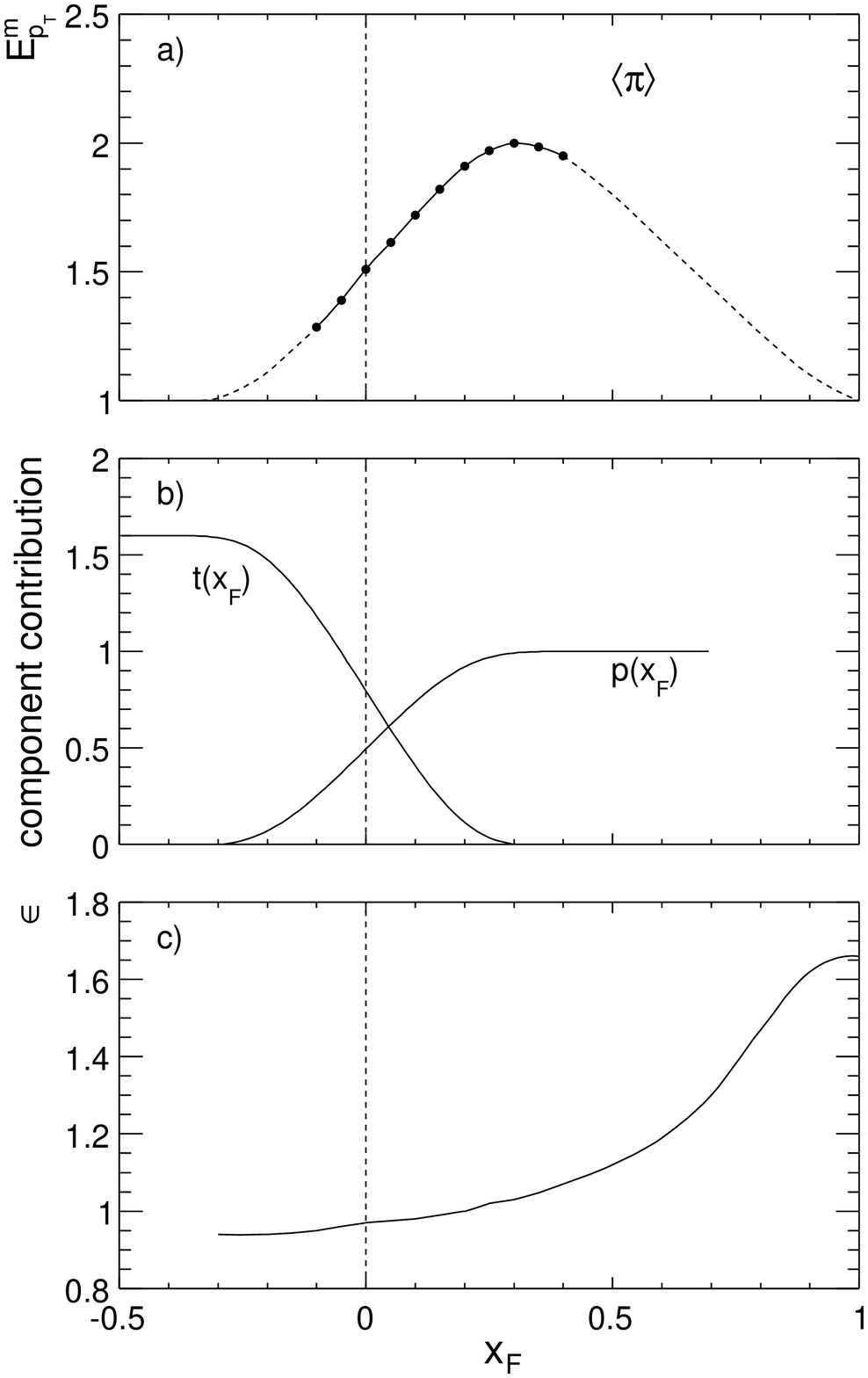,width=8.8cm}
  \caption{a) The enhancement factor $E^m_{p_T}$ as a function of $x_F$, 
           b) target and projectile contributions as a function of $x_F$ 
	   at $p_T$~=~1.8~GeV/c, and c) efficiency function $\epsilon$ 
	   multiplying the relative contribution from single collisions 
	   as a function of $x_F$ }
  \label{fig:targ_sch}
\end{figure}
  
In a two-component superposition scheme, the first phenomenon is
indeed expected in the negative $x_F$ limit where the projectile
component vanishes with respect to the target contribution. In
this picture there is no reason to believe that the fragmentation
of the participant target nucleons shows anomalous behaviour at high $p_T$.
In other words the target enhancement factor $E_{p_T}^\textrm{target}$ 
should be equal to one over the full $x_F$ range. The superposition 
should be characterized by target and projectile components with a 
specific $x_F$ dependence, as presented in Fig.~\ref{fig:targ_sch}b.

The second phenomenon is of different origin. It is connected to
the large probability of single collisions, $P(1)$, in minimum bias
proton-nucleus interactions, as discussed in Sects.~\ref{sec:pp_two} and 
\ref{sec:mult_coll} above. It has been shown there that the contribution 
from single collisions grows as $x_F$~=~1 is approached since the pion 
yield from multiple collisions is progressively suppressed at large 
$x_F$. As $E^m_{p_T}$ is unity for single collisions, the enhancement should 
tend towards one for large $x_F$. The influence of $P(1)$ on the variation 
of the measured enhancement with $x_F$ can be described by a multiplicative 
weight function $\epsilon(x_F)$ as shown in Fig.~\ref{fig:targ_sch}c. 
This function approaches the value of 1.6~$\sim$~1/$P(1)$ for 
$x_F \rightarrow$~1 and reduces to $\sim$~0.9 in the backward region due to 
the yield increase from the projectile fragmentation.

%
%
\subsection{Extraction of the projectile component}
\vspace{3mm}
 
Concerning the target-projectile superposition, it can be stated
that the range of the corresponding target-projectile feed-over is
larger than at low transverse momentum (Sect.~\ref{sec:pc_two}) and reaches 
$x_F \sim \pm$0.3. This indicates an important $p_T$ dependence. This is 
quantified by writing down the composition of the measured enhancement 
$E^m_{p_T}$ from the target ($E^\textrm{target}_{p_T}$~=~1) 
and the projectile ($E^\textrm{proj}_{p_T}$) components:

\begin{equation} 
  E^m_{p_T} = \frac{t(x_F)E^\textrm{target}_{p_T} + 
     r_\textrm{proj} E^\textrm{proj}_{p_T}}{t(x_F) + r_\textrm{proj}}
\label{eq:meas_enh}
\end{equation}
where $r_\textrm{proj} = (dn/dx_F)^\textrm{pC}_\textrm{proj}(\langle p_T \rangle)/
             (dn/dx_F)^\textrm{pp}(\langle p_T \rangle)$.
Here $t(x_F)$ and $p(x_F)$ specify the shape of the relative target and
projectile contributions as a function of $x_F$. This shape is constrained
by the boundary conditions $p$(-0.3)~=~$t$(0.3)~=~0 and $p$($x_F >$~0.3)~=~1,
$t$($x_F <$~-0.3)~=~$\langle \nu \rangle$, and the symmetry conditions  
$p$(0)~=~0.5 and $t$(0)~=~$\langle \nu \rangle$/2 as indicated in 
Fig.~\ref{fig:targ_sch}. Evidently these functions are more similar to
the proton than to the low-$p_T$ pion feed-over shown in Fig.~\ref{fig:pipr_fover}.
The enhancement factor for the projectile component can be extracted
from Eq.~\ref{eq:meas_enh} as

\begin{equation} 
  E^\textrm{proj}_{p_T} = \frac{1}{r_\textrm{proj}} \left( E^m_{p_T} (t(x_F)
          + r_\textrm{proj}) - t(x_F)\right)
\end{equation}
 
\begin{figure}[h]
  \centering
  \epsfig{file=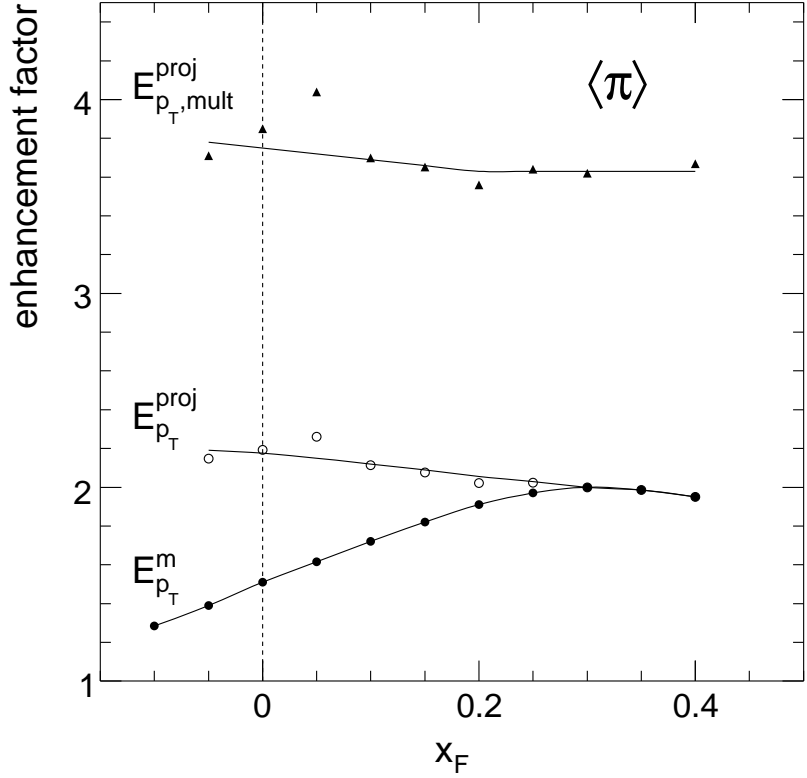,width=11.0cm}
  \caption{The enhancement factors $E^m_{p_T}$, $E^\textrm{proj}_{p_T}$, 
           $E^\textrm{proj}_{p_T,\nu \geq 2}$ as a function of $x_F$ at 
           $p_T$~=~1.8~GeV/c }
  \label{fig:proj_enh}
\end{figure}

The $x_F$ dependence of $E^\textrm{proj}_{p_T}$ is presented in 
Fig.~\ref{fig:proj_enh}.

As an important consequence of this decomposition it appears that the
measured nuclear enhancement at $x_F$~=~0 is not fully representative of
the phenomenon. It is very substantially increased if proper account
is taken of the ``inert'' target contribution in p+A reactions. This
is very important for the understanding of the nuclear enhancement
in the symmetric A+A interactions where the target and projectile
contribution are equally contributing. From this consideration
alone it must be expected that the measured nuclear enhancement in
A+A collisions must be larger than in p+A collisions. If however
properly extracting the projectile component the enhancement should
tend to be equal in both types of reaction at equal $\langle \nu \rangle$ 
per projectile (p+A) or participant (A+A), unless new physics phenomena 
should be present in A+A interactions.

%
%
\subsection{Enhancement due to multiple collisions}
\vspace{3mm}
  
The influence of single collisions on the measured nuclear enhancement
is particularly important in the p+C case discussed here, as the
probability $P(1)$ is about 0.6 and the corresponding enhancement 
should be unity for this contribution. Taking account of the  
weight function $\epsilon(x_F)$ introduced above, the enhancement 
due to multiple collisions can be extracted as

\begin{equation} 
  E^\textrm{proj}_{p_T,mult}(x_F) = 
  \frac{E^\textrm{proj}_{p_T}(x_F) - P(1) \epsilon(x_F)}
       {1-P(1) \epsilon(x_F)}
  \label{eq:corr_enh}
\end{equation}
 
The application of Eq.~\ref{eq:corr_enh} results in a further, very 
considerable increase of the nuclear enhancement over the full $x_F$ 
range as evident from Fig.~\ref{fig:proj_enh}. The final result for 
$E^\textrm{proj}_{p_T,mult}$ is very different indeed 
from the measured quantity $E^m_{p_T}$. In this context
the application of straight-forward parametrizations of the
A-dependence, as they are applied in the description of the nuclear
enhancement, Sect.~\ref{sec:nuc_enh} above, is to be regarded with great 
caution. This because the probability $P(1)$ depends strongly on A, 
see Fig.~\ref{fig:nu_prob}. By relating the effect in p+Be to the one 
in p+W for instance, one compares cases where $P(1)$ differs by a more 
than a factor of 3. This means that the exponent $\alpha(p_T)$ will 
decrease appreciably once the trivial effect of single collisions is 
taken out.

As far as A+A collisions are concerned, the correct way of comparison
would be to relate minimum bias p+A to minimum bias A+A collisions
where the distributions $P(\nu)$ are equal \cite{bib:andrzej}. 
For centrality selected collisions, this condition is more difficult to 
fulfill as the probability $P(1)$ stays always greater than zero for 
A+A interactions even for impact parameter zero, whereas it vanishes 
for central p+A interactions \cite{bib:andrzej}.
 
%
%
\subsection{A look at rapidity dependence}
\vspace{3mm}
 
\begin{figure}[h]
  \centering
  \epsfig{file=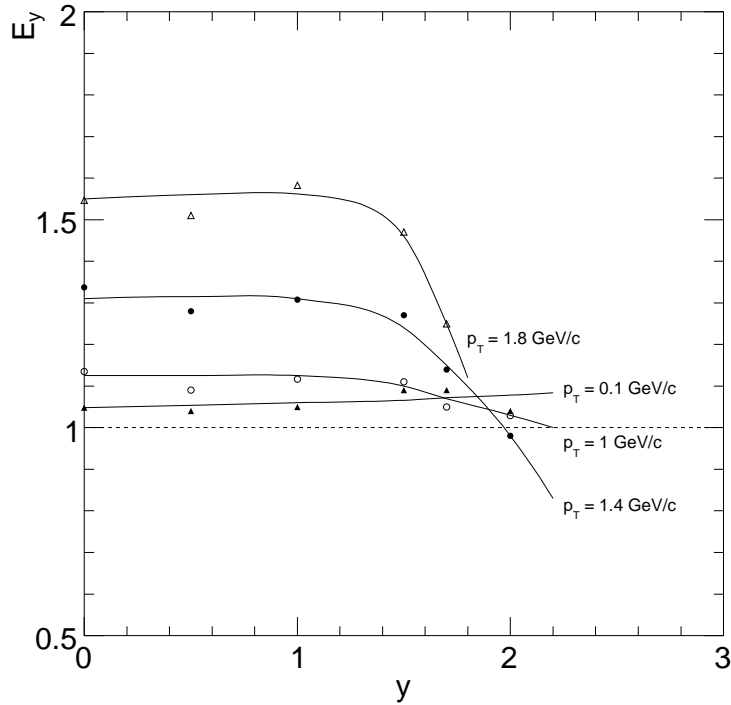,width=10cm}
  \caption{The $p_T$ enhancement as a function of $y$ at several fixed $p_T$ values }
  \label{fig:rap}
\end{figure}

The interpretation of the nuclear enhancement as a $p_T$ dependent
phenomenon necessitates the use of kinematic variables that are
orthogonal in the transverse and longitudinal components, as is
the case for the $x_F$/$p_T$ combination used here. By describing the
same phenomenon in the variables rapidity $y$ and $p_T$, a completely
different picture will emerge as the use of $y$ combines both
the transverse and the longitudinal evolution of the particle
densities, except, of course, at $y$~=~0. This is exemplified in 
Fig.~\ref{fig:rap} where the $y$ dependence of the enhancement factor 

\begin{equation} 
  E_y(y,p_T) = \frac{R_y(y,p_T)}{R_y(y,\langle p_T \rangle)} 
\end{equation}
with $R_y(y,p_T) = (dn/dx_F)^\textrm{pC}(y,p_T)/(dn/dx_F)^\textrm{pp}(y,p_T)$,
is presented at the same four values of transverse momentum as shown
in Fig.~\ref{fig:ept}.

The strong increase of the nuclear enhancement with $x_F$ is replaced
by a flat $y$-dependence up to $y \sim$~1, followed by a strong decrease
for $y >$~1.5. From this plot alone one might erroneously conclude
that the Cronin effect vanishes in forward direction, whereas is
reality one is looking at the longitudinal decrease of particle
density in p+A collisions, see Sect.~\ref{sec:pc_two} above.
 
Experimentally it is often difficult, if not impossible, to fulfill
the orthogonality condition mentioned above. This is true for
all experiments being limited by a polar angle (or rapidity) cutoff
in forward direction, where the low $p_T$ region may not be accessible at
all above a certain limit of $x_F$. This is typical of collider experiments.
For the BRAHMS detector at RHIC, for instance, the rapidity limit of 3.2 units
corresponds to a $p_T$ cutoff at 0.8~GeV/c already at $x_F$~=~0.1 where
the measured nuclear enhancement is still increasing, and at
$p_T$~=~2 GeV/c for $x_F$~=~0.25 where it reaches its maximum.
 
%
%
\section{The trace of resonance decay}
\vspace{3mm}

The production and decay of resonances constitutes a much neglected
but important field of studies concerning soft hadronization. As
already pointed out in \cite{bib:pp_paper} many details of the inclusive 
pion distributions directly show the imprint of resonance decay. In the
context of the present paper several of the extracted features
concerning the two-component nature of the hadronization process
will be discussed here in relation to resonance decay. This concerns
in particular the following experimental findings:

\begin{itemize}
\item The shape and feed-over of the extracted, $p_T$ integrated projectile
      component (Sects.~\ref{sec:pp_two} and \ref{sec:pc_two}).
\item Its dependence on particle mass (Sect.~\ref{sec:baryon}).
\item Its important $p_T$ dependence up to the limit of the NA49 data at
      1.8~GeV/c (Sect.~\ref{sec:high_pt}).
\end{itemize}      
                                                                            
\begin{figure}[h]
  \centering
  \epsfig{file=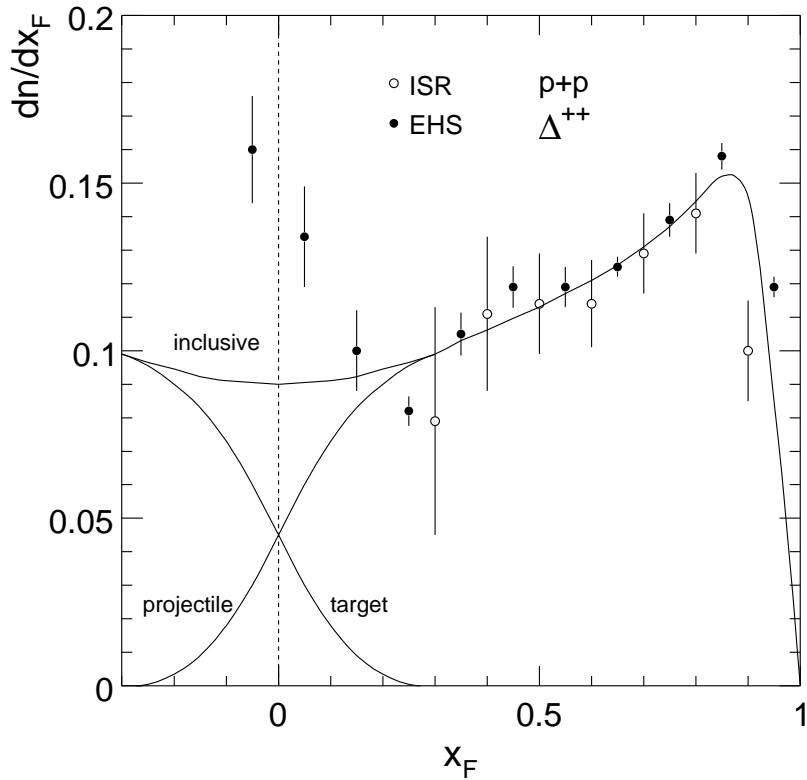,width=11cm}
  \caption{Two-component picture for $\Delta^{++}$}
  \label{fig:delta}
\end{figure}

\begin{figure}[t]
  \centering
  \epsfig{file=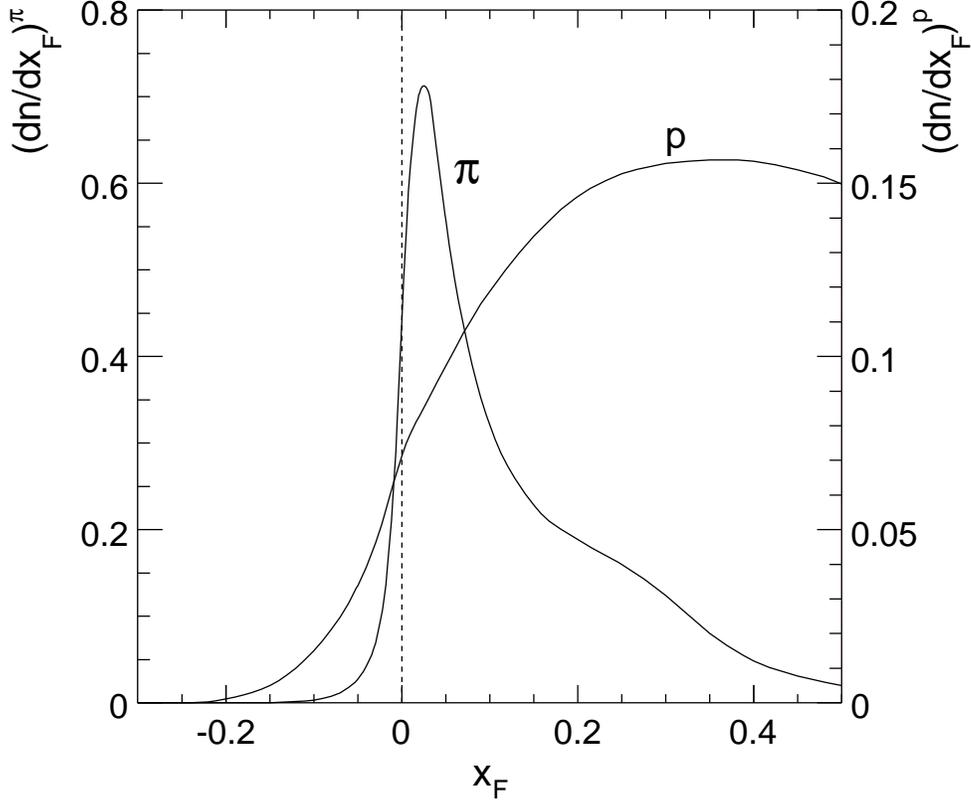,width=13cm}
  \caption{$p_T$ integrated $x_F$ distributions of protons and pions 
           produced from $\Delta^{++}$ decay }
  \label{fig:dec_prod}
\end{figure}

As a complete treatment of the manifestations of resonance phenomena
exceeds the scope of this paper and has to be left to a coming publication,
only some basic mechanisms which are of straight-forward kinematic origin
will be treated here, using a single resonance as a tool for explanation.
The resonance chosen is the $\Delta^{++}$(1232) whose inclusive $x_F$ and 
$p_T$ distributions have been determined in an adequate fashion for the
present purpose \cite{bib:delta}. Its measured, $p_T$ integrated $x_F$ 
distribution is shown in Fig.~\ref{fig:delta} together with the target and 
projectile components constructed from it, using the baryonic feed-over 
shape as discussed in Sect.~\ref{sec:baryon}.                                                                                  
The two-body decay $\Delta^{++} \rightarrow \textrm{p}+\pi^+$ performed 
for the projectile component of Fig.~\ref{fig:delta} yields $p_T$ 
integrated $x_F$ distributions of the decay products as shown in 
Fig.~\ref{fig:dec_prod}.
                              
A first observation concerns the overall shape of the $\pi^+$ distribution
in comparison to the extracted inclusive projectile contributions shown
in Figs.~\ref{fig:pp_2comp} and \ref{fig:shape}. Although there are of course 
deviations in detail, as to be expected from the single, low-mass resonance 
studied here, the overall agreement in shape and $x_F$-range is very evident. 
This concerns specifically the limited range of the pion feed-over into the
target hemisphere.
 
A second observation concerns the difference of this feed-over between
proton and pion. The mass dependence quantified in Fig.11b is reproduced
insofar as the baryon distribution reaches much further into the opposite
hemisphere than for the pion.
 
This effect is of course a straight-forward consequence of the Lorentz
transformation between the resonance rest frame and the overall cms.
The energy factor of this transformation reads
 
\begin{equation}
  E = \sqrt{m_\textrm{sec}^2 + q^2}
  \label{eq:mass}
\end{equation}
where $m_\textrm{sec}$ is the decay particle mass and $q$ is given by

\begin{equation} 
  q = \frac{1}{2}m_\textrm{res}\sqrt{(m_\textrm{res}^2 - 
    (m_\pi + m_p)^2)(m_\textrm{res}^2 - (m_\pi - m_p)^2)}.
  \label{eq:q}
\end{equation}
 
At the nominal mass of the $\Delta^{++}$ this corresponds to 0.229~GeV/c
which is much smaller than the proton mass. The proton therefore
receives, in contrast to the pion, a major fraction of the resonance
momentum in the cms of the interaction.
 
\begin{figure}[t]
  \centering
  \epsfig{file=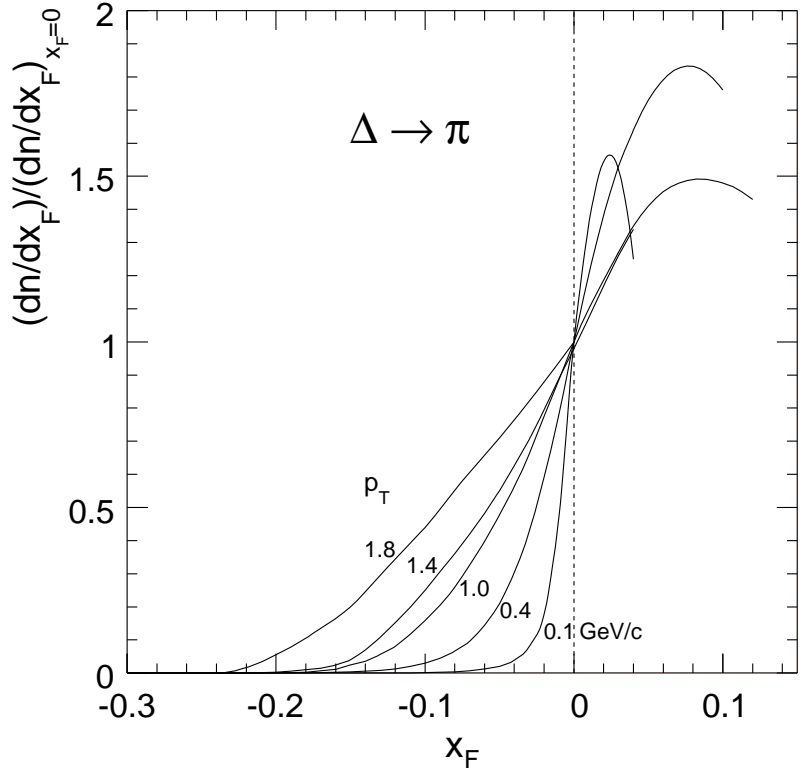,width=11cm}
  \caption{Pion feed-over from $\Delta^{++}$ decay for different values of $p_T$}
  \label{fig:delta_pt}
\end{figure}

A third observation concerning $p_T$ dependence is obtained by studying
the $\Delta^{++}$ decay for pions in bins of transverse momentum, as shown
in Fig.~\ref{fig:delta_pt}.

For the same bins used in Fig.~\ref{fig:ept} above, the distributions are 
normalized to unity at $x_F$~=~0 in order to calibrate the strong $p_T$ 
dependence of the cross section. Evidently there is a steady increase of 
the range of feed-over with increasing transverse momentum. The shape 
dependence of the $p_T$ integrated $x_F$ distributions on the lower 
integration limit demonstrated in Figs.~\ref{fig:c2p_ptcut} and 
\ref{fig:c2p_charge} finds thereby a natural explanation. At the upper 
$p_T$ limit of the NA49 data at 1.8~GeV/c, the shape of the pion distribution 
approaches the one measured for the proton. This effect has been used for the
extraction of the projectile component of the $p_T$ enhancement, 
Fig.~\ref{fig:targ_sch}.
 
This third observation follows again from the resonance decay kinematics.
In fact the most efficient way to obtain high transverse momentum
for low mass secondaries is to exploit the Breit-Wigner tail of the resonance
mass distribution which reaches, in the case of the $\Delta$(1232), values
well above 2~GeV. Correspondingly the decay momentum $q$ (Eq.~\ref{eq:q}) 
increases into the 1~GeV/c region which in turn reduces the mass 
dependence (Eq.~\ref{eq:mass}) and increases the range of the feed-over. 
It is therefore necessary to take proper account of the resonance mass 
distribution in any quantitative work on hadronization dealing with 
resonance decay.
 
Following these observations, resonance decay as a source of ``high $p_T$''
production in the sense defined above has to be considered very seriously.
Using a sum of 13 measured resonances it has indeed been demonstrated
for p+p collisions \cite{bib:hgf2} that the inclusive pion distributions resulting
from two body decays saturate the yields up to $p_T$~=~3~GeV/c and beyond
simultaneously over the full range of $x_F$. A modification of the
resonance spectrum towards higher masses in multiple collisions would
therefore be sufficient to also describe the anomalous nuclear
enhancement in this $p_T$ range, at SPS energies, without invoking
partonic or perturbative effects.

%
%
\clearpage
\section{Conclusions}
\vspace{3mm}

This paper attempts a new assessment of the p+A interaction with light
nuclei at SPS energy which is based exclusively on experimental data.
The study leads to a number of conclusions which cover a wide range of
observables. They may be subsummed as follows:

\begin{itemize}
\item[$\bullet$]The reference to a set of experimental results obtained in 
      elementary and nuclear interactions, in particular to the new precision 
      data from the NA49 detector in p+C and p+p collisions, allows for a model
      independent argumentation.
\item[$\bullet$] This argumentation clearly establishes the
       composition of the p+A reaction from three basic components:
       projectile fragmentation, target fragmentation, and
       intranuclear cascading.
\item[$\bullet$] These contributions have been isolated and quantified one by one.
\item[$\bullet$] Intranuclear cascading governs the far backward hemisphere. Its
      contribution rapidly decreases towards $x_F$~=~0 and corresponds to 
      only a couple of percent relative to the participant fragmentation at
      $x_F$~=~-0.1.
\item[$\bullet$] For the remaining target and projectile contributions, a clear two
      component picture has been established. The two components overlap
      in a limited zone around $x_F$~=~0. The extent of this zone is shown to
      depend on particle species and transverse momentum. It is limited
      to $¦x_F¦ \pm$~0.1 for the $p_T$ integrated pion yield and reaches out
      to $¦x_F¦ \pm$~0.25 at $p_T$ of 1.8~GeV/c.
\item[$\bullet$] Target fragmentation reflects the superposition of participant
      hadronization, revealing the detailed charge dependence introduced
      by the isospin composition of the nucleus. The total pion yield
      from this source corresponds to an average of 1.6 participant
      target nucleons which is close to but significantly below the number
      expected from measured nuclear parameters.
\item[$\bullet$] The extraction of detailed features of the projectile fragmentation
      has been the main aim of the study. In relation to elementary collisions
      a complex pattern of deviations emerges which concerns practically all
      measured quantities.
\item[$\bullet$] The total pion yield increases by about 10\% as compared to 
      the elementary interaction. This increase is concentrated near $x_F$~=~0 
      with a shape that matches the density distribution of the projectile component.
\item[$\bullet$] Above $x_F \sim$~0.2 the differential yield decreases steadily 
      with respect to elementary collisions as $x_F$ increases, with a specific 
      charge dependence.
\item[$\bullet$] Using the well defined probability for single projectile collisions,
      the contribution from multiple collisions has been extracted and
      compared to the centrality controlled p+C data subsample.
\item[$\bullet$] The detailed $p_T$ dependence of the projectile component has been
      established in comparison to elementary collisions. In the neighbourhood
      of the mean transverse momentum the comparison is characterized by
      important, charge dependent local structure. With increasing
      transverse momentum a general increase of pion yield is established.
\item[$\bullet$] This increase is quantified by an enhancement factor which shows
      strong $x_F$ dependence and has been corrected for the inert contributions
      from target fragmentation and from single collisions.
\item[$\bullet$] The relation to the anomalous nuclear enhancement (Cronin effect) is
      established and it is demonstrated that a new assessment of this
      phenomenon is mandatory from a purely experimental point of view.
\item[$\bullet$] Finally the importance of resonance production and decay for 
      virtually all studied quantities is evoked.
\end{itemize}
                                                                                  
This study will be continued and deepened by using the data on centrality
selected p+Pb and Pb+Pb collisions that are also available from the
NA49 experiment.

\section*{Acknowledgements}
\vspace{3mm}
This work was supported by
the Polish State Committee for Scientific Research 
(1 P03B 006 30, SPB/CERN/P-03/Dz 446/2002-2004, 2 P03B 04123),
the Hungarian Scientific Research Foundation (T032648, T032293, T043514),
the Hungarian National Science Foundation, OTKA, (F034707),
the Polish-German Foundation,
the Bulgarian National Science Fund (Ph-09/05),
the EU FP6 HRM Marie Curie Intra-European Fellowship Program,
and 
the Particle Physics and Astronomy Research Council (PPARC) of the United Kingdom.


\clearpage

\begin{thebibliography}{[20]}
\bibitem{bib:hgf} H.~G.~Fischer et al., Nucl. Phys. {\bf A715} (2003) 118
\bibitem{bib:pp_paper} C.~Alt et al., Eur. Phys. J. {\bf C45} (2006) 343
\bibitem{bib:pc_paper} Inclusive production of charged pions in p+C collisions at 
                       158 GeV/c beam momentum, hep-ex/0606028 (2006) (to be published in Eur. Phys. J.)
\bibitem{bib:num_coll} A.~Bialas et al., Nucl. Phys. {\bf B111} (1976) 461
\bibitem{bib:barton} D.~Barton et al., Phys. Rev. {\bf D27} (1983) 2580
\bibitem{bib:hgf1} H.~G.~Fischer et al., Heavy Ion Phys. {\bf 17} (2003) 369
\bibitem{bib:bobbink} G.~J.~Bobbink et al., Nucl. Phys. {\bf B204} (1982) 173
\bibitem{bib:bromberg} C.~M.~Bromberg et al., Phys. Rev. {\bf D9} (1974) 1864
\bibitem{bib:kafka} T.~Kafka et al., Phys. Rev. Lett. {\bf 34} (1975) 687
\bibitem{bib:aivazyan} V.~V.~Aivazyan et al., Z. Phys. {\bf C42} (1989) 533
\bibitem{bib:uhlig} S.~Uhlig et al., Nucl. Phys. {\bf B132} (1978) 15
\bibitem{bib:guettler} K.~Guettler et al., Nucl. Phys. {\bf B116} (1976) 77
\bibitem{bib:field} R.~D.~Field and R.~P.~Feynman, Nucl. Phys. {\bf B136} (1978) 1
\bibitem{bib:dezso} D.~Varga, Study of Inclusive and Correlated Particle Production
                    in Elementary Hadronic Interactions, PhD. Thesis (2003) 
                    E\"otv\"os Lor\'and University, Budapest
\bibitem{bib:dezso1} D.~Varga et al., Heavy Ion Phys. {\bf 17} (2003) 387
\bibitem{bib:andrzej} A.~Rybicki, H.Niewodnicza\'nski Institute of Nuclear Physics, 
                      Internal Report No 1976/PH, 
		      http://www.ifj.edu.pl/publ/reports/2006/
\bibitem{bib:el_scat} E.~A.~J.~M.~Offermann et al., Phys. Rev. {\bf C44} (1991) 1096 \\
                      I.~Sick, Phys. Lett. {\bf 116B} (1982) 212
\bibitem{bib:book} H.~Enge, Introduction to Nuclear Physics, Addison-Wesley World Student Series (1972)
\bibitem{bib:fricke} G.~Fricke et al., Atomic Data and Nuclear Data Tables {\bf 60} 
                     (1995) 177
\bibitem{bib:sick} I.~Sick, Nucl. Phys. {\bf A218} (1974) 509
\bibitem{bib:hh1} M.~K.~Hegab and J.~H\"ufner, Nucl. Phys. {\bf A384} (1982) 353
\bibitem{bib:bialas} A.~Bialas et al., Phys. Lett. {\bf B51} (1974) 179      
\bibitem{bib:niki} N.~A.~Nikiforov et al., Phys. Rev. {\bf C22} (1980) 700
\bibitem{bib:coulomb} J.~Burfening et al., Phys. Rev. {\bf 75} (1949) 382 \\
                      H.~Yagoda, Phys. Rev. {\bf 85} (1952) 891 \\
                      E.~M.~Friedl\"ander, Phys. Lett. {\bf 2} (1962) 38 \\
                      N.~I.~Kostanashvili et al., Sov. J. Nucl. Phys. {\bf 13} (1971) 715
\bibitem{bib:abduz} A.~Abduzhamilov et al., Phys. Rev. {\bf D39} (1989) 86
\bibitem{bib:nim} S.~Afanasiev et al., Nucl. Instrum. Meth. {\bf A430} (1999) 210 
\bibitem{bib:negra} M.~Della~Negra et al., Nucl. Phys. {\bf B127} (1977) 1 \\
                    R.~P.~Feynman et al., Phys. Rev. {\bf D18} (1978) 3320
\bibitem{bib:gaard} J.~J.~Gaardh\o je et al., Eur. Phys. J. {\bf C43} (2005) 287 \\
                    B.~Z.~Kopeliovich et al., Phys. Rev. {\bf C72} 054606
\bibitem{bib:cronin} D.~Antreasyan et al., Phys. Rev. {\bf D19} (1979) 764
\bibitem{bib:cron_par} J.~K\"uhn, Phys. Rev. {\bf D13} (1976) 2948 \\
                       A.~Krzywicki et al., Phys. Lett. {\bf B85} (1979) 407 \\
		       M.~Lev and B.~Petersson, Z. Phys. {\bf C21} (1983) 155 \\
		       A.~Accardi, hep-ph/0212148 (2003) \\
		       B.~Kopeliovich et al., Phys. Rev. Lett. {\bf 88} (2002) 232303
\bibitem{bib:delta} A.~Breakstone et al., Z. Phys. {\bf C21} (1984) 321  \\
                    M.~Aguilar-Benitez et al., Z. Phys. {\bf C50} (1991) 405
\bibitem{bib:hgf2} H.~G.~Fischer et al., CERN/SPSC 2005-035
\end{thebibliography}
\end{document}